\documentclass{aa}  

\newcommand{\final}[1]{\textcolor{black}{#1}} % Final corrections
\newcommand{\oliver}[1]{\textcolor{black}{#1}} % Oliver Voelkel
\newcommand{\revision}[1]{{#1}} % Referee Report Comments

\usepackage{graphicx}
\usepackage{txfonts}
\usepackage{natbib}
\bibpunct{(}{)}{;}{a}{}{,} % to follow the A&A style
\usepackage{hyperref}
\hypersetup{colorlinks=true, urlcolor=blue, linkcolor=blue, citecolor=blue}
\usepackage{comment}
\usepackage{amstext}

\begin{document}

\title{On the multiple generations of planetary embryos}
\titlerunning{On the multiple generations of planetary embryos}

\author{
    Oliver Voelkel\inst{\ref{mpia}}
    \and
    Hubert Klahr\inst{\ref{mpia}}
    \and
    Christoph Mordasini\inst{\ref{unibe}}
    \and 
    {Alexandre Emsenhuber\inst{\ref{lmu}}}
}
\authorrunning{O. Voelkel et al.}

\institute{
    Max Planck Institute for Astronomy, Heidelberg, Königstuhl 17, 69117 Heidelberg, Germany \\
    \email{voelkel@mpia.de} \label{mpia}
    \and
    Physikalisches Institut, University of Bern, Gesellschaftsstrasse 6 CH 3012 Bern, Switzerland 
    \label{unibe}
    \and 
    Universitäts-Sternwarte München, Ludwig-Maximilians-Universität München, Scheinerstraße 1, 81679 München, Germany
    \label{lmu}
    %\and
    %Lunar and Planetary Laboratory, University of Arizona, 1629 E. University Blvd., Tucson, AZ 85721, USA \label{lpl}
}
    
%\date{Received DD MM YYYY / Accepted DD MM YYYY}

% \abstract{}{}{}{}{} 
% 5 {} token are mandatory
\abstract
  % context heading (optional)
  % {} leave it empty if necessary  
   {
    Global models of planet formation tend to begin with an initial set of planetary embryos for the sake simplicity. While this approach gives valuable insights on the evolution of the initial embryos, the initial distribution itself is a bold assumption. Limiting oneself to an initial distribution may neglect essential physics that precedes, or follows said initial distribution.
   }
  % aims heading (mandatory)
   {
    We wish to investigate the effect of dynamic planetary embryo formation on the formation of planetary systems.
   }
  % methods heading (mandatory)
   {
    The presented framework begins with an initial disk of gas, dust and pebbles. The disk evolution, the formation of planetesimals and the formation of planetary embryos is modeled consistently. Embryos then grow by pebble, planetesimal and eventually gas accretion. Planet disk interactions and N-body dynamics with other simultaneously growing embryos is included in the framework. 
   }
  % Results heading (mandatory) 
   {
    We show that the formation of planets can occur in multiple consecutive phases. Earlier generations grow massive by pebble accretion but are subject to fast type I migration and thus accretion to the star. The later generations of embryos that form grow to much smaller masses by planetesimal accretion, as the amount of pebbles in the disk has vanished.
   }
  % conclusions heading (optional), leave it empty if necessary 
   {
    The formation history of planetary systems may be far more complex than an initial distribution of embryos could reflect. The dynamic formation of planetary embryos needs to be considered in global models of planet formation to allow for a complete picture of the systems evolution.
   }


\maketitle
%
%________________________________________________________________
%
\section{Introduction}
\subsection{Motivation}
\oliver{
Latest observational constraints on grain growth in young protostellar disks via thermal dust emission \citep{harsono2018evidence} imply that the formation of planets may begin in the earliest embedded phases in the life of young stars. The idea that previous generations of gas giant planets may have been accreted by the host star as a result of inward migration has already been introduced by \cite{lin1986tidal}. This hypothesis states that the final system of planets that can be observed around a star may only reflect a small subset of the planets that initially formed. The possibility of a previous protogiant planet in the solar system is mentioned as well, however not the existence of previous super Earths or other terrestrial mass planets.
}
The number of planets that form and survive during the lifetime of a circumstellar disk is unknown. Only the minimum number of survived planets per system is a lower constraint, as it is given as the number of exoplanet detections. While this number often lacks completeness due to the low detectability of low mass planets in certain systems, it completely lacks the information on the previous history of the system.
\final{
Already discussed is also that planet bearing stars might be polluted and show higher metalicities \citep{gonzalez1997stellar,murray2002stars}.
}
It is thus possible that the currently observed population of planets merely reflects a small fraction of the planets that initially formed.
While the research conducted on how individual planets grow and evolve, based on an initially placed embryo continues to flourish, the initial formation of the used embryo is typically an initial assumption. 
\final{
As recently shown in \cite{schlecker2021new}, this initial embryo location is the initial condition with the highest predictive power on the the outcome of a planet.
}
The question on how many embryos form and how many of those survive however cannot be neglected if one attempts to study the formation of planets in a consistent fashion.
Essential to this is a model that predicts the number of planetary embryos from the previously evolved system and tracks their combined evolution until the dispersal of the circumstellar disk.
This study will present a self consistent global model of planet formation and disk evolution that allows for such a study. The presented model enables us to investigate the number of planets that form and evolve within a circumstellar disk over its entire lifetime. We find that the formation history of planetary systems is far more complex than an initial distribution of planetary embryos could reflect.
\subsection{Global models of planet formation}
\label{subsec:globalModelsOfPlanetFormation}
To clarify the terminology, we wish to give a brief overview on current models of planet formation, focusing on their differences, similarities and limitations. One similarity that all planet formation models bring with themselves is that the approach aims to combine multiple physical processes in a common framework. This results from the complexity of the problem, as the formation of planets cannot be described in an isolated manner. Not only ranges the process from a dust grain to a gas giant over numerous orders of magnitude in mass, it also needs to be embedded in the global evolution of a circumstellar disk. In addition to the evolution of the disk itself , interactions with other simultaneously growing planets can influence planet formation. Planet-planet interaction, as well as planet-disk interactions can decide the fate of a planet during its formation and/or later evolutionary stages. 
 
Current global models of planet formation focus either on a specific time in planet formation, a specific accretion mechanism (pebble accretion or planetesimal accretion), or a specific location in the disk. Models that have been introduced by \cite{ida2004toward}, \cite{alibert2005models}, \cite{mordasini2012combined} and \cite{emsenhuber2020new} focus on the accretion of planetesimals on initially placed planetary cores. Even though the size of these planetesimals is its own ongoing field of research, here we refer to planetesimals as objects in the size range from 600$\,$m in diameter \citep{emsenhuber2020new} to 100$\,$km in diameter \cite{voelkel2020effect}. 
\oliver
{
Their size plays a major role in the accretion mechanism, as more massive objects are less likely to be accreted and the stirring by a protoplanet increases their eccentricities and inclinations. Even at sizes of several hundred meters to km however, the gas disk can significantly damp the planetesimal dynamical states and make them a highly efficient mechanism for protoplanetary growth \citep{emsenhuber2020new}.
}
\oliver
{
The accretion of smaller particles for which gas drag can cause the object to even spiral onto the accreted protoplanet is called pebble accretion \citep{ormel2010effect}.
}
Planet formation models that are built around the accretion of pebbles onto initially placed planetary embryos have been introduced in \cite{bitsch2015growth}, \cite{Ndugu_2017} and \cite{brugger2020pebbles}. The aforementioned planet formation models either focus on the accretion of planetesimals or the accretion of pebbles. Hybrid accretion models have recently been introduced by e.g. \cite{alibert2018formation} or \cite{guilera2020giant}. A major drawback of these models however, as well as models that study pebble or planetesimal accretion in an isolated fashion is the initial placement of planetary embryos. This initial assumptions skips the earliest phase of the circumstellar disk evolution in which the planetesimals form that later accumulate to planetary embryos. As results from \cite{bitsch2015growth} show, the location and the time when an embryo is placed plays a dominant role in the subsequent evolution. Recent work presented in \cite{voelkel2021linkingI} and \cite{voelkel2021linkingII} studies the formation of planetary embryos from planetesimals that form from an evolving pebble disk. They find that more distant embryos (>2-3$\,$au) form after the pebble flux has largely vanished. While the accretion of pebbles on planetesimals and planetary embryos is included in \cite{voelkel2021linkingII}, only the innermost planetary embryos can benefit from pebble accretion, as the outer embryos fail to form during the lifetime of the pebble flux. 
 
Recent work by \cite{guilera2020giant} studied the formation of giant planets around pressure bumps. They combined a global disk evolution model containing gas, dust and pebble dynamics with the formation of planetesimals due to streaming instability behind the pressure bumps. The embryo was placed once the mass in planetesimals is equivalent of the mass of a planetary embryo. They also used a hybrid accretion model that combines pebble and planetesimal accretion, as well as a global disk evolution model. While in  their extensive model they did not use the embryo as an initial assumption, the embryo was placed in a specific location, which was subject to an initial assumption as well. The formation time of the embryo in their work is given by the time it takes until a lunar mass of 100$\,$km planetesimals has formed around their pressure bump. While this is a first constraint on the initial placement time, it does not account for planetesimal growth by planetesimal collisions, which can take significantly longer than it takes to form the 100$\,$km planetesimals themselves. Additionally, their placed embryo was the only embryo in the system, which neglects planet planet interactions. While this model attempts to model all stages of the single planet in the system (beginning from dust and pebbles to a final gas giant) using a global disk evolution model, the planet formation studied remains local, as the location and the number of planets in the system remains fixed.
 
A planet formation model based on planetesimal accretion that forms planetesimals consistent with the radial evolution of dust and pebbles is also presented in \cite{voelkel2020effect}. This approach connected a two population model for dust and pebble dynamics \citep{birnstiel2012simple} with the pebble flux regulated model for planetesimal formation from \cite{Lenz_2019}. The evolution of dust, pebbles and planetesimals was merged with the planet formation model from \cite{emsenhuber2020new} to study the impact of different planetesimal distributions on planet formation. While planetesimals formed consistently with the disk evolution, pebble accretion was neglected and planetary embryos remained an initial assumption. The formation model of \cite{emsenhuber2020new} however is capable of also tracking the growth and N-body dynamics/interactions of up to 100 planetary embryos. A model that forms planetary embryos based on the local planetesimal surface density evolution is presented and discussed in \cite{voelkel2021linkingI} and \cite{voelkel2021linkingII}. To bridge the gap between disk evolution and the accretion of pebbles and planetesimals onto planetary embryos, we decided to implement the embryo formation model from \cite{voelkel2021linkingI} into the planet formation model described in \cite{voelkel2020effect} and model the accretion of both pebbles and planetesimals. 
 
We wish to highlight here that the planet formation model presented in this work combines the currently known accretion mechanisms (pebble accretion, planetesimal accretion, gas accretion) during the entire lifetime of the circumstellar disk, that includes the formation of planetary embryos, as well as late stage gas accretion. The number, as well as the formation time and location of planetary embryos are no longer an assumption, but the result of the analytic embryo formation model from \cite{voelkel2021linkingI}. 
\oliver
{
This stands in contrast to the models of \cite{emsenhuber2020newpop} where a fixed number of embryos (1,20,50,100) is inserted at t=0 throughout the disk uniformly in log.
}
In the following we will discuss the individual stages of the planet formation model used in this study. A detailed description of the different stages of planet formation that are covered in the model that is presented in this paper are described in Sect. \ref{Sec:Our_global_model}.
\section{Our global model of planet formation}
\label{Sec:Our_global_model}
In the following we will discuss the different stages of planet formation that are covered in our global formation model. As a reminder, the computation of the disk evolution, the accretion of material, the formation of embryos and planetary migration are computed simultaneously. The existence of a planet changes the pebble flux due to pebble accretion. The formation of planetesimals (as it is regulated by the pebble flux) changes accordingly and thus the formation of other planetary embryos is affected as well. 
\subsection{Disk evolution model}
\label{subsec:diskEvolModel}
A detailed description on the implementation of the 
\oliver
{
gas and solid disk 
}
evolution model used in our planet formation framework can be found in \cite{voelkel2020effect}, here we will discuss the underlying fundamentals. We use a one dimensional disk evolution model that tracks the evolution of gas, dust, pebbles and planetesimals. The viscously evolving gas disk \citep{lust1952, lyndenbellpringle1974} uses an $\alpha$-prescription for turbulence \citep{shakura1973black}
\oliver
{
and includes internal and external photoevaporation \citep{picogna2019dispersal}.
}
\revision{The inner edge of the disk is modeled as in \cite{alibert2005models}, setting the surface density at the boundary to the floor value of the gas and the mass flux through the inner boundary equal to the equilibrium flux.}
Coupled to the evolution of the gas disk is the two population solid evolution by \cite{birnstiel2012simple}. This model solves an advection diffusion equation of a combined solid density. Depending on whether the particles at a given radial location can be considered in the drift or in the fragmentation limit of growth, a fixed mass relation is applied. This relation splits the solid density into two populations. Depending on the individual Stokes number of the particles, these populations can be considered as dust (St$<<1$) or pebbles (St$\ge1$) respectively. 
 
The formation of planetesimals in our framework is regulated by the local radial pebble flux. The model we use has been introduced by \cite{Lenz_2019} and it does not specify which physical mechanism (e.g. Kelvin Helmholtz instability or streaming instability) drives the formation of planetesimals. Its underlying assumption is that planetesimals form in trapping zones that can appear at any location of the disk. These trapping zones appear for a given lifetime and with a radial separation of $d(r)$. In \cite{Lenz_2019} planetesimals form proportional to the radial pebble flux and the formation rate of planetesimals is given as
\begin{align}
    \dot{\Sigma}_p(r) = \frac{\epsilon}{d(r)} \frac{\dot{M}_{peb}(r)}{2 \pi r}
\end{align}
with $\Sigma_P(r)$ as the local planetesimal surface density at a heliocentric distance $r$ and $\dot{M}_{peb}(r)$ as the local radial pebble flux. $\epsilon$ describes the amount of the pebble flux that is transformed into planetesimals over the trap distance $d(r)$. The distance of pebble traps is given as 5 gas pressure scale heights in our approach \revision{ as in \cite{Lenz_2019}. The reason for this assumption stems from the typical separation of zonal flows found in \cite{dittrich2013gravoturbulent} }. Planetesimals are assumed to be in the oligarchic regime \citep{idamakino1993,thommes2003,chambers2006}. 
\oliver{
They are described in a fluid type fashion using a surface density $\Sigma_P$ with eccentricity and inclination. For their dynamical state we use the approach from \cite{fortier2013}. 
Planetesimals are stirred by the embryos, as well as by each other and damped by the gas disk. This stirring by the protoplanet follows \cite{guilera2010consequences} while the planetesimal planetesimal stirring follows \cite{ohtsuki2002evolution}. The damping of planetesimals follows \cite{inaba2001high} in the quadratic regime, \cite{adachi1976gas} and \cite{rafikov2004fast} in the Stokes and Epstein regime.
}
The size at which planetesimals form is given as 100$\,$km in diameter. While the size of planetesimals is an ongoing field of research, we choose a size of 100$\,$km in diameter, as observational constraints from the solar systems infer \citep{bottke2005linking,walsh2017identification,Delbo2017} and what numerical simulations suggest \citep{schafer2017initial, Klahr2020}. Other works suggest smaller sizes, in the range of several 100$\,$m to kilometres in diameter \citep{arimatsu2019kilometre, schlichting2013initial, weidenschilling2011initial, zheng2017planetesimal}.
\subsection{Planetesimals to planetary embryos}
\label{subsec:embryo_formation}
Planetesimals are described as a one dimensional surface density ($\Sigma_p$), analogous to gas, dust and pebbles. While we do not track the N-body evolution of this large number of planetesimals, we track the dynamical N-body evolution of up to 100 planetary embryos. These embryos are introduced over time into the simulation, consistent with the evolution of the planetesimal surface density and its dynamical state. The embryo formation model that we use has been introduced by \cite{voelkel2021linkingI} and will be briefly described in the following. 
 
Once planetesimals begin to form, we track their growth by integrating the local mass growth rate of a planetesimal in the oligarchic regime within a swarm of planetesimals \citep{lissauer1993planet}
\begin{align}
    \frac{d M_{p}(r,t)}{d t} 
    = 
    \frac{\sqrt{3}}{2} \Sigma_P(r,t) \cdot \Omega(r) \pi r{_b}^2 
    \left(
    1 + \frac{v^2_{\text{esc}}(M_p,R_b)}{v^2_{\infty}(r,t)}
    \right)
    \label{eq:mass_growth}
\end{align}
with $M_{p}(r,t)$ as the mass of the largest object at a heliocentric distance $r$ at a time $t$. $\Sigma_P (r,t)$ is given as the local planetesimal surface density, $\Omega(r)$ as the orbital Kepler frequency, $R_{b}$ the radius of the largest object, $v_{\text{esc}}$ as the escape velocity of $M_p$ at its surface and $v_{\infty}$ as the dispersion velocity of planetesimals, which we give as $v_{\infty} = e(r) \cdot \Omega(r)$ with $e(r)$ as the planetesimals eccentricity.
$M_{p}$ is initially set to the mass of a 100$\,$km planetesimal with a solid density of $\rho_s = 1.0\,$g/cm$^3$
\begin{align}
    M_{p}(r,t_{0}) = M_{100\,km}
\end{align}
Once $M_{p}$ locally surpasses the mass of a planetary embryo, which in our case is given as a lunar mass ($M_{emb} = 0.0123$M$_{\oplus}$), a new N-body object is introduced into the simulation and $M_p$ is reset to $M_{p}(r,t) = M_{100\,km}$ within 15 Hill radii of the placed embryo. An additional constraint from \cite{voelkel2021linkingI} is that an embryo cannot form within 15R$_{Hill}$ of any other embryo in the system. The orbital separation of planetary embryos in the oligarchic growth regime has already been found and confirmed in \cite{kokubo1998oligarchic,Kobayashi_2011,walsh2019planetesimals} and serves as a good constraint on the number of planetary embryos within a given spatial distribution. While the model has been derived without including the effect of pebble accretion, the effect of pebble accretion on embryo formation has been studied in \cite{voelkel2021linkingII} in a similar framework, including pebble accretion. It is shown that the accretion of pebbles largely affects the mass growth rate of planetary embryos \oliver{ with masses >0.01 M$_{\oplus}$} and thus their physical spacing to each other. When expressed in the embryo Hill radii however, the orbital separations remain similar to what has been found in \cite{voelkel2021linkingI,kokubo1998oligarchic,Kobayashi_2011,walsh2019planetesimals}.
\oliver{
Pebble accretion only begins to be an effective accretion mechanism at masses much larger than that of a 100km planetesimal. The initial planetesimal growth from 100$\,$km to a lunar mass object therefore remains dominated by planetesimal collisions. The initial formation time of a lunar mass object is therefore only weakly influenced by pebble accretion \citep{voelkel2021linkingII}. Its subsequent growth however begins to be dominated by pebble accretion. Including the embryo formation model from \cite{voelkel2021linkingI} into our planet formation framework therefore ensures that the total number of embryos in the system, their formation time and their spatial distribution is consistent with the evolution of the planetesimal surface density and its dynamical state. 
\revision{The initial eccentricity of an embryo is chosen randomly between $e=10^{-3}$ and $e=10^{-5}$. The initial inclination is given as $i=e/2$.}
}
\subsection{Embryos and beyond}
Once a planetary embryo has formed according to Sect. \ref{subsec:embryo_formation} it is subject to several simultaneously occurring processes. Its mass growth is given by the accretion of pebbles, planetesimals and gas \citep{pollack1996formation}. To avoid confusion, there is no physical difference in our model between a planetary embryo and a planet. The terminology of an embryo only refers to the object at its initial lunar mass of 0.0123$\,$M$_{\oplus}$. It is then treated as a single N-body object and will be referred to as planet in the following. Every planet that formed in our model was initially introduced to the systems as a lunar mass embryo based on the model described in Sec. \ref{subsec:embryo_formation}. During its growth, the planet is subject to planetary migration and the dynamical interaction with other planets in the system. Planets can be scattered out of the system \revision{if they reach an orbital distance to the host star of above 1000$\,$au}. \revision{Mergers with other simultaneously forming planets are included as well, if the orbital distance between two planets is less than the sum of their radii. The density of the core is computed as in \cite{mordasini2012characterization} by using a modified polytropic EOS from \cite{seager2007mass}. } 
 
As mentioned, planetesimals are considered to be in the oligarchic regime. They are accreted by embryos and evolve their eccentricity and inclination by self stirring, by the interaction with the embryo and by the damping of the gas disk (see Sec. \ref{subsec:diskEvolModel}). Next to the accretion of planetesimals, we included the accretion of pebbles from the disk, based on the prescription of \cite{ormel2017emerging}. The accretion of pebbles is considered until the planet reaches its local pebble isolation mass \citep{lambrechts2014separating}. Once planets are large enough to accrete gas from their surrounding, the 1D structure of the gas envelope is retrieved by solving the internal structure equations \citep{bodenheimerpollack1986}. The accretion of solids affects the accretion of gas during the initial phase via accretional heating \citep{pollack1996formation,leechiang2015,alibert2018formation}. As seen in \cite{voelkel2020effect}, this effect can suppress runaway gas accretion for high planetesimal surface densities, as runaway gas accretion can only occur if the accretion rate of gas surpasses the accretion rate of planetesimals \citep{pollack1996formation} . 
 
As planets that are embedded in the gas disk grow, they can undergo type I and type II migration. Type I migration is treated as described in \cite{coleman2014formation}, type II migration as in \cite{dittkrist2014} and to distinguish between them we use the prescription by \cite{crida2006} for gap opening. \oliver{
As discussed in \cite{coleman2014formation}, the formation of strong corotation torques can also lead to type I outward migration within our framework.
}
\revision{
The total type I torque exerted onto the planet is given as \citep{paardekooper2011torque,coleman2014formation}
\begin{align}
\label{eq:total_torque}
    \Gamma_{1} = F_{L}\Gamma_{L} + F_{e} F_{i} (\Gamma_{\text{c,baro}} + \Gamma_{\text{c,ent}})
\end{align}
with $\Gamma_{L}$ as the Lindblad torque and $\Gamma_{c,baro}$ and $\Gamma_{c,ent}$ as the barotropic and entropic corotation torque. The reduction of the Lindblad torque caused by the eccentricity and inclination of the planets orbit follows \cite{cresswell2008three} and is given as 
\begin{equation}
        F_{L}^{-1} = P_{e} + \left( \frac{P_{e}}{|P_{e}|} \right) 
        (0.07 \hat{i} + 0.085 \hat{i}^{4} - 0.08 \hat{e} \hat{i}^{2} ) 
\end{equation}
with 
\begin{equation}
    P_{e} = \frac{1 + (\frac{\hat{e}}{2.25})^{1/2} + (\frac{\hat{e}}{2.84})^{6} }
    {1 - (\frac{\hat{e}}{2.02})^4}
\end{equation}
using $\hat{e} = e/h_{asp}$ and $\hat{i} = i/h_{asp}$ with $h_{asp} = h/r$ as the aspect ratio of the disk.
$F_e$ and $F_i$ in Eq. \ref{eq:total_torque} denote the reduction of the corotation torque as caused by the planets eccentricity and inclination \citep{bitsch2010orbital}.
They are given as 
\begin{eqnarray}
    \label{eq:F_e}
    F_e &=& \exp{\left( - \frac{e}{h_{asp}/2 + 0.01} \right)} \\
    F_i &=& 1 - \tanh{\hat{i}}
    \label{eq:F_i}
\end{eqnarray}
see \cite{fendyke2014corotation} for Eq. \ref{eq:F_e} and \cite{coleman2014formation} for Eq. \ref{eq:F_i}. The damping of the planets eccentricity and inclination is included as in \cite{cresswell2008three}.
}
\revision{
The vertical structure of gas disk is computed at each time step of the disks evolution using the approach of \cite{nakamoto1994formation}. The vertical scale height is then given as $h = c_s / \Omega$ with $c_s = \sqrt{k_b T_{mid} / (\mu m_H)}$ (using $\mu = 2.24$ as molecular weight and $m_H$ as the hydrogen atom mass). The midplane temperature $T_{mid}$ is given as 
\begin{equation}
    \sigma_{SB} T_{mid}^{4} = 
    \frac{1}{2} \left(
    \frac{3}{8} \tau_R
    +
    \frac{1}{2 \tau_P}
    \right)
    \dot{E}
    +
    \sigma_{SB} T_{s}^{4}
\end{equation}
with $T_{s}$ as the irradiation temperature, $\sigma_{SB}$ as the Stefan-Boltzmann constant, $\tau_{R}$ and $\tau_{P}$ as the Rosseland and Planck mean optical depths and $\dot{E}$ as the viscous dissipation rate.
}
The entire framework is described in great detail in \cite{emsenhuber2020new}, whereas here we only give a brief overview over the included physics. Strictly speaking, \cite{emsenhuber2020new} refer to a model for planet population synthesis, that contains a planet formation model at its heart. In this sense \cite{emsenhuber2020new} is an updated version of \cite{mordasini2018planetary}, in which a model for planet formation \citep{alibert2005models,alibert2013theoretical} and planet evolution \citep{mordasini2012combined} is combined to carry out a planet population synthesis approach, as in \citet{mordasini2009extrasolar}. As we will not carry out a population synthesis in this study, but investigate a single system, we will not make use of the population synthesis capabilities of our used framework. However we wish to mention here, that the entire framework presented in this paper is capable of conducting the same population synthesis studies as presented in e.g. \cite{emsenhuber2020newpop} or \cite{schlecker2021new}, as our additional physical models (pebble and dust dynamics, pebble accretion,  planetesimal formation and planetary embryo formation) do not require high computational costs.
\section{Numerical setup and initial conditions}
\label{Sec:numerical_setup}
We focus on a set of disk parameters that has been introduced in \cite{lenz2020constraining} (see Table \ref{tab:initial_parameters}). Their attempt was to constrain the parameters of the solar nebula by conducting a large parameter study. The parameters found resulted in what \cite{lenz2020constraining} refer to as the "most appealing solar nebula" (MASN). The distribution of planetesimals that resulted from the aforementioned parameters did best in constraining the solar nebula, based on the distribution of planets and asteroids in the solar system today. 
%
%In our fourth simulation, we let $\alpha(r,t)$ evolve as a function of the disks temperature. The evolution of $\alpha$ is chosen so that we find the same jump in %the turbulence as in Eq. \ref{eq:alpha_jump_fixed} around the evolving iceline. $\alpha(r,t)$ is then given as 
%\begin{align}
%    \alpha(r,t) = \left\{\begin{array}{ll} 
%        3 \times 10^{-4}, & r < r_{ice}(t) - 0.25\, \text{au} \\
%        1 \times 10^{-4}, & r > r_{ice}(t) + 0,25\, \text{au}
%        \end{array}\right.
%        \label{eq:alpha_jump_iceline}
%\end{align} .
\begin{table}
    \centering
    \begin{tabular}{l l l}
    \hline
    \hline 
    Symbol                  &  Value                          & Meaning                             \\
    \hline    \\
    M$_\text{star}$         &  1.0$\,$M$_\odot$              & Mass of the central star             \\
    
    M$_\text{disk}$         &  0.1$\,$M$_\odot$              & Total mass of the gas disk           \\
                  
    a$_\text{in}$           &  0.03$\,$au                    & Inner disk radius                    \\
      
    a$_\text{out}$          &  20$\,$au                      & Exponential cutoff radius            \\
    
    $\gamma$                &  1.0                           & initial $\Sigma_g$ profile ($\Sigma_g \propto r^{-\gamma}$)  \\
                  
    $d_g$                   &  1.34 $\times 10^{-2}$         & Dust-to-gas ratio                    \\
                                        
    $\alpha$                &  3.0 $\times 10^{-4}$          & Turbulence parameter                 \\
                                        
    L$_\text{x}$            &  3.0 $\times 10^{29}\,$ergs/s  & X-ray luminosity                    \\
                                            
    v$_\text{frag}$         &  200$\,$cm/s                   & Fragmentation velocity  \\
                                            
    $\epsilon$              &  0.05                          & $\Sigma_P$ formation efficiency \\
    
    $d$                     &  5 h                           & $\Sigma_P$ trap distance \\
    
    $h$                     &  $c_s / \Omega$                & gas pressure scale height \\
    
    $\tau_f$                &  1600 t$_{\text{orbit}}$       & $\Sigma_P$ trap lifetime \\
                                            
    $\rho_s$                &  1.0 g/cm$^{3}$                & $\Sigma_P$ solid density \\
    \hline                       
    \\
    \end{tabular}
    \caption{ Disk and planetesimal formation parameters used in our study. The set of parameters stems from the most appealing solar nebula, as described in \cite{lenz2020constraining}. \oliver{The planetesimal trap distance $d$ is set to 5 gas pressure scale heights} \revision{($h=c_s/\Omega$ with $c_s$ as the local speed of sound and $\Omega$ as the Kepler frequency.)}
    }
    \label{tab:initial_parameters}
\end{table}
\section{Simulation results}
We investigate the effect of dynamic embryo formation on the formation of a planetary system. While this extensive model allows for a multitude of effects to investigate in greater detail, our study focuses on the composition, mass and semimajor axis evolution of the resulting planetary system (Sec. \ref{subsubsec:system_evolution}), the number of active and formed planets over time (Sec. \ref{subsec:numberOfPlanetsOverTime}), the evolution of the disk surface densities and masses (Sec. \ref{subsec:diskEvolution}), the evolution of the solid mass components (Sec. \ref{subsec:solidMassEvolution}) and the final system of planets (Sec. \ref{subsec:finalSystem}). 
\subsection{Planetary system evolution}
\label{subsubsec:system_evolution}
\begin{figure}[]
\centering
\includegraphics[width=1.0\linewidth]{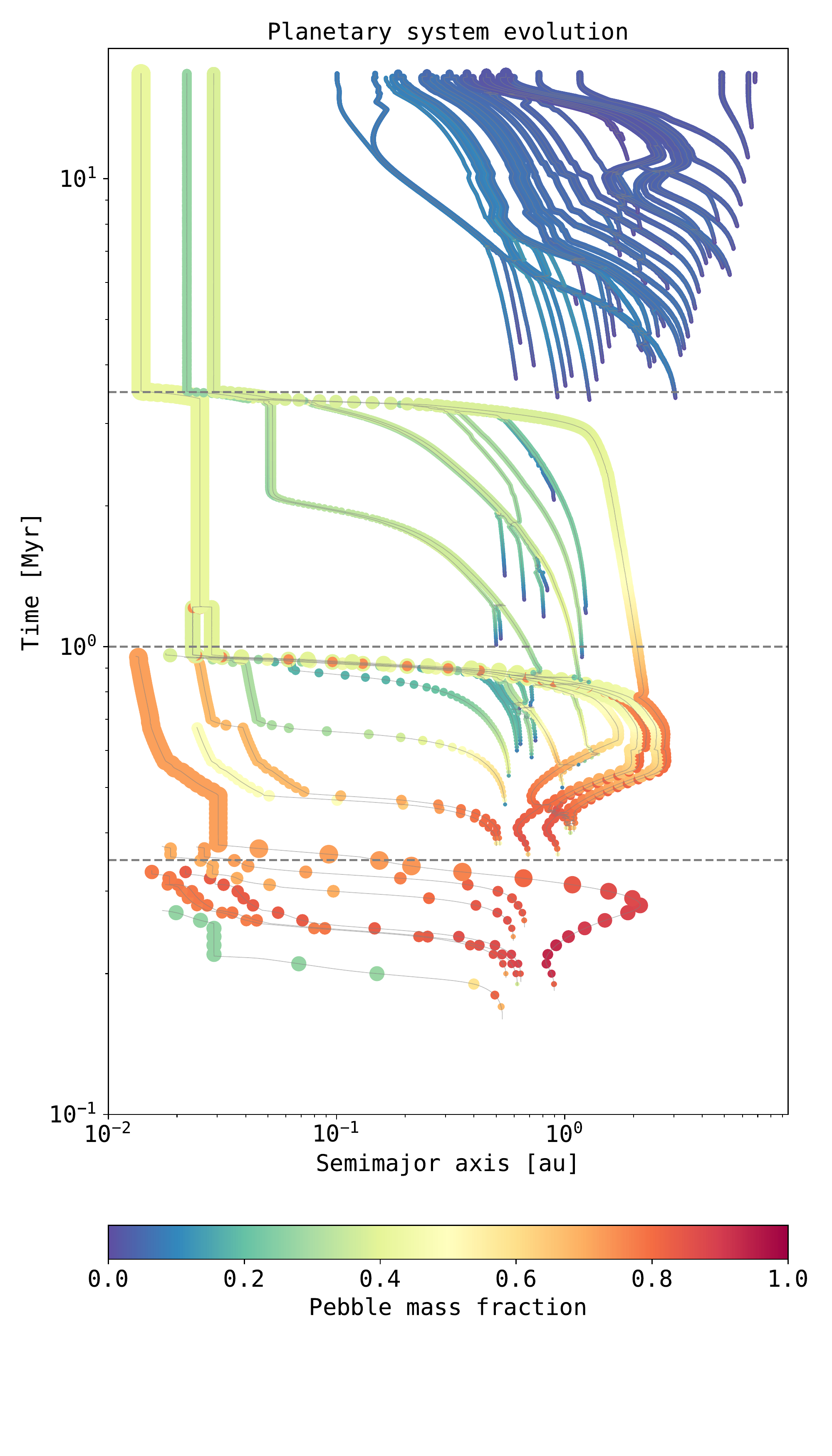}
\caption{\small  Semimajor axis over time evolution for the planetary system. The evolution of a planet is linked via the grey line and the size of the dots indicate the mass of the planets every 10$\,$ky. The color of the planets shows their pebble mass fraction M$_{\text{peb}}$/M$_{P}$. The horizontal lines are drawn at 0.35$\,$Myr, 1$\,$Myr and 3.5$\,$Myr and show the moments at which most currently active planets are accreted by the host star or were subject to mergers. \oliver{Thus we find four distinct generations of planet formation.}
} 
\label{Fig:semimajorAxisTime_3e-4}
\end{figure}

\begin{figure}[]
\centering
\includegraphics[width=1.0\linewidth]{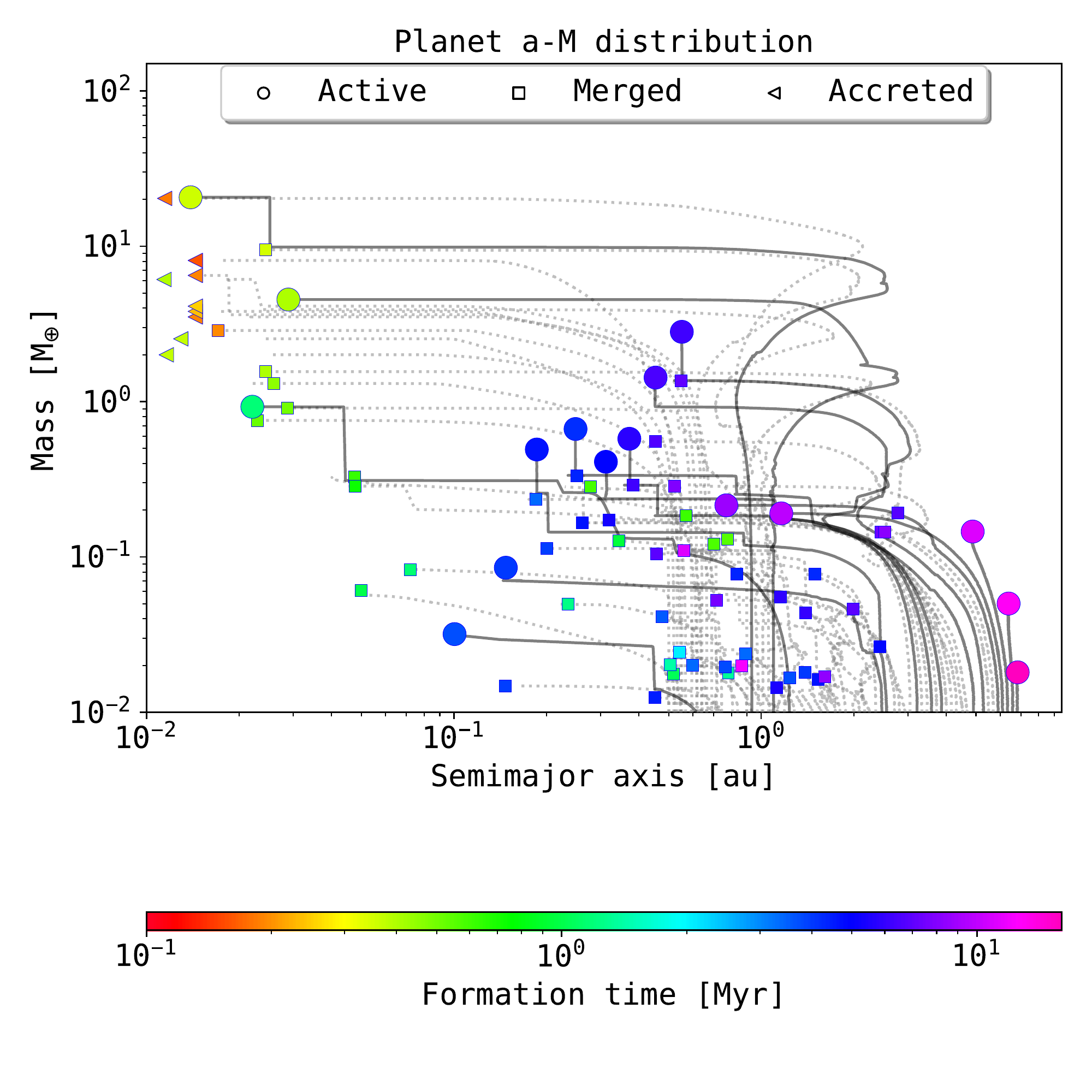}
\caption{\small  \oliver{
Mass and semimajor axis distribution of the entire planetary system until the dispersal of the gas disk. The large circle markers indicate planets that remain active until the end of the simulation. Planets that were accreted by the host star are shown as triangles and planets that merged via collisions with other planets are indicated as squares. The track of the planets is shown as the solid grey line for active planets and dotted grey lines for accreted or merged planets. The color of the final marker indicates the initial formation time of the corresponding planetary embryo. 
}
} 
\label{Fig:a_M_distribution}
\end{figure}

\begin{figure}[]
\centering
\includegraphics[width=1.0\linewidth]{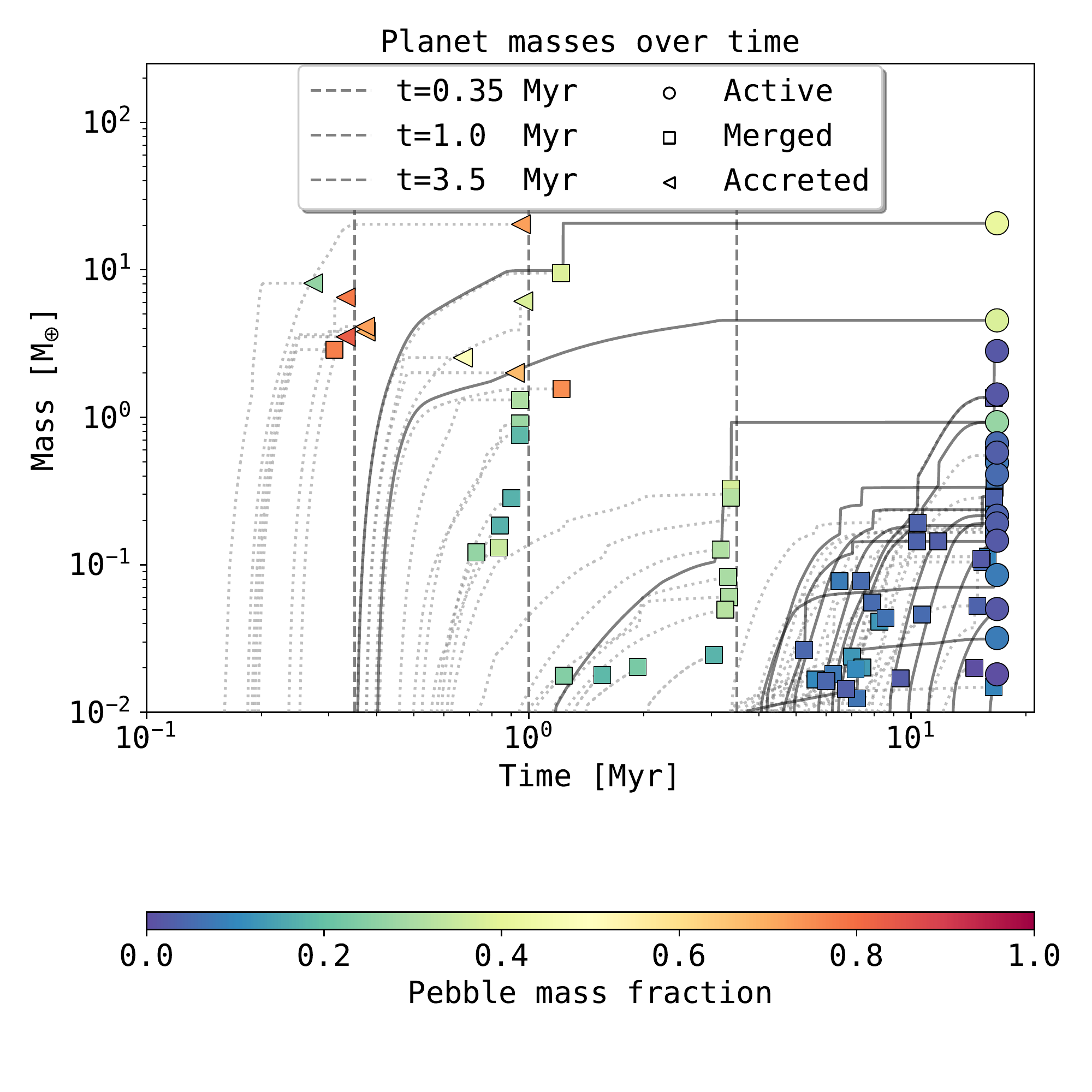}
\caption{\small  Planet mass over time. The vertical lines are drawn at 0.35$\,$Myr, 1 Myr and 3.5 Myr and show the moments at which most currently active  planets are accreted by the host star or were subject to mergers. \oliver{The vertical steps are caused by giant impacts that result in planet mergers. The large circle markers indicate planets that remain active until the end of the simulation. Planets that were accreted by the host star are shown as triangles and planets that merged via collisions with other planets are indicated as squares. The track of the planets is shown as the solid grey line for active planets and dotted grey lines for accreted or merged planets. The color of the final marker indicates the pebble mass fraction of the corresponding planet. }
} 
\label{Fig:planetMass}
\end{figure}

Fig. \ref{Fig:semimajorAxisTime_3e-4} shows the mass, pebble mass fraction and semi-major axis over time of all planets during the lifetime of the gas disk. \oliver{ Fig. \ref{Fig:a_M_distribution} shows the mass and semimajor axis distribution of the system, including their growth tracks and the initial formation time of the corresponding planetary embryo.}
Fig. \ref{Fig:planetMass} shows the evolution of the corresponding planet masses. In Fig. \ref{Fig:semimajorAxisTime_3e-4}, within the first 0.35$\,$Myr, we find that several embryos form within 1$\,$au and grow dominantely by pebble accretion, as given by the color of the dots. As it can be seen in Fig. \ref{Fig:planetMass}, the most massive planet reaches up to 20$\,$M$_{\oplus}$. These early formed planets experience strong type I migration and eventually end at the inner edge of the gas disk. \oliver{As more planets migrate to the inner edge, the innermost planets are accreted by the host star, as shown by the triangular markers in Fig. \ref{Fig:a_M_distribution}.} The same evolution is underwent by the next set of planets that forms after 0.35$\,$Myr. Within the first 1$\,$Myr, we see that several new embryos form in the terrestrial planet zone. Initially dominated by pebble accretion, those super Earth mass planets first experience outward migration \oliver{ due to positive corotation torques \citep{coleman2014formation},} followed by inward migration before halting at the inner edge of the gas disk again. Those planets form within 1.2$\,$au but temporarily reach a semimajor axis of 2-3$\,$au. Within the first 3.5$\,$Myr, we see that one planet does not migrate to the inner edge of the disk within the first 1$\,$Myr, but over a significantly longer timescale. As it can be seen by the pebble mass fraction evolution of the outermost planet at 1$\,$Myr, it was initially dominated by pebble accretion. Over the course of the next 2.5$\,$Myr, its pebble mass fraction strongly decreases due to ongoing planetesimal accretion. During that phase, a set of sub Earth mass planets has formed in the area around 1$\,$au. After the outer super Earth goes into another phase of type I inward migration, it pushes the sub Earth mass planets to the inner edge of the gas disk as well, eventually clearing the terrestrial planet zone of planets. \revision{The migration of planets beyond the innner edge of the gas disk is due to N-body interactions with other planets. This involves resonant chains and scattering up to the point of reaching the surface of the host star.} After the last super Earth has migrated to the inner edge of the gas disk at 3.5$\,$Myr, a large set of sub Earth mass planets emerges over the next 13$\,$Myr. Those planets stay at smaller masses than their super Earth predecessors for the lifetime of the gas disk and consequentially experience significantly slower type I migration. As it can be seen in Fig. \ref{Fig:planetMass}, the two most massive planets of the remaining system have formed at 0.35$\,$Myr as part of the second generation of embryo formation and no planet from the first generation of embryo formation survived until the dispersal of the gas disk. 
\subsection{Number of planets over time}
\label{subsec:numberOfPlanetsOverTime}
\begin{figure}[]
\centering
\includegraphics[width=1.0\linewidth]{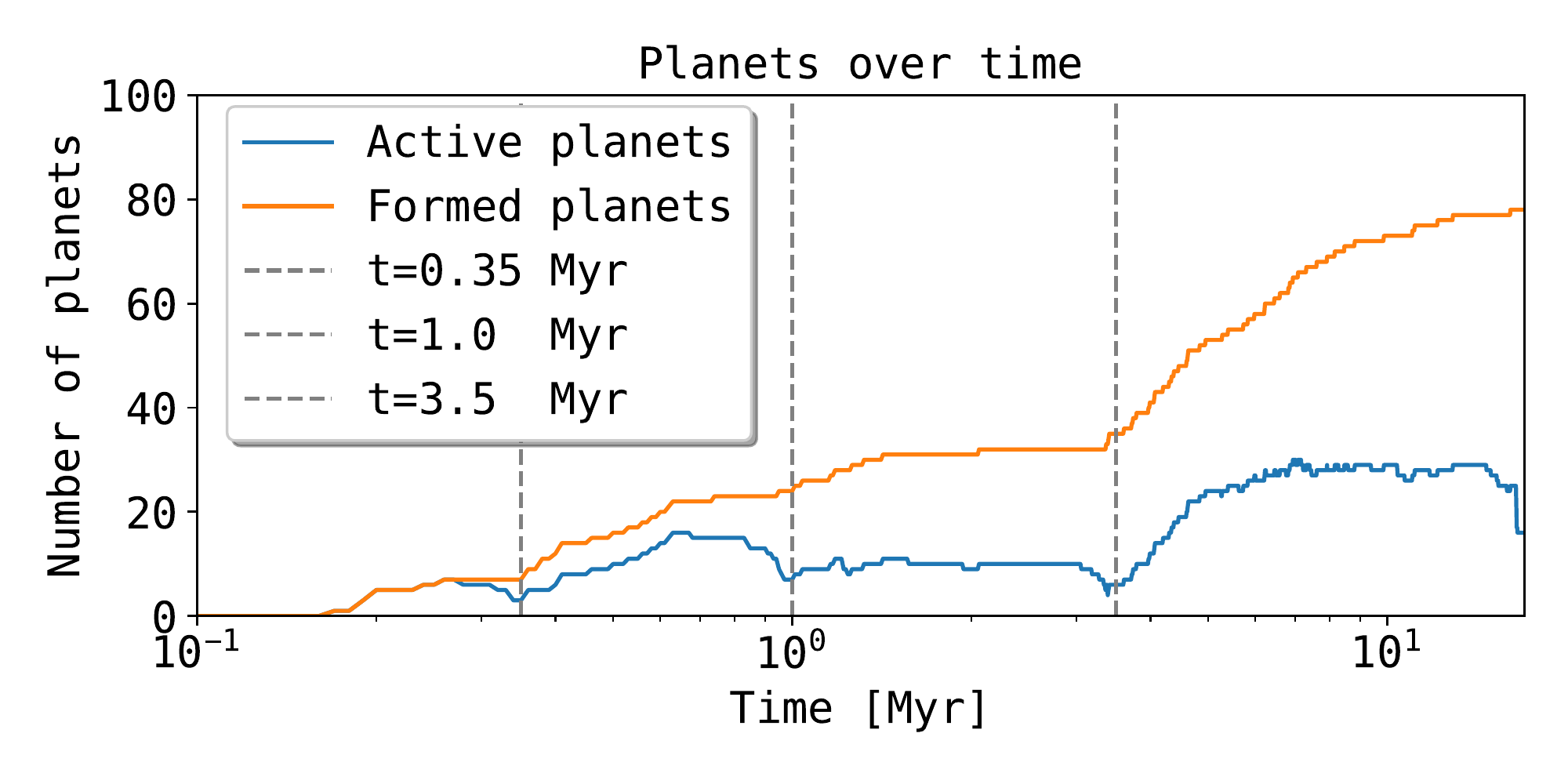} \\
\caption{\small  Number of planets over time. The blue line shows the number of currently active planets in the system, while the orange line shows the total number of planets that have formed. The vertical lines are drawn at 0.35$\,$Myr, 1$\,$Myr and 3.5$\,$Myr and show the moments at which most currently active  planets are accreted by the host star or were subject to mergers.
} 
\label{Fig:numberOfPlanets}
\end{figure}

In Fig. \ref{Fig:numberOfPlanets} we show the total number of active planets and the cumulative number of planets that formed during the lifetime of the gas disk. As the number of formed planets continuously increases, the number of active planets shows three significant moments of decrease. The first decrease can be found at 0.35$\,$Myr, the second at 1$\,$Myr and the third at 3.5$\,$Myr. The local minima of active planets in the disk are indicated by the vertical lines. The total number of planets that formed during the lifetime of the disk is given as 78, whereas the total number of active planets after the lifetime of the disk is given as 16. The largest number of planets formed in the last generation after 3.5$\,$Myr. \oliver{ While the number of formed planets keeps increasing after 5 Myr, the number of active planets after that time remains almost constant at $\sim$30 due to giant impacts and mergers. In the latest stages after 16 Myr, the number of active planets drops to 16, as planets continue to collide and merge, but no more embryos are forming.}
\subsection{Disk evolution}
\label{subsec:diskEvolution}
\begin{figure}[]
\centering
\includegraphics[width=1.0\linewidth]{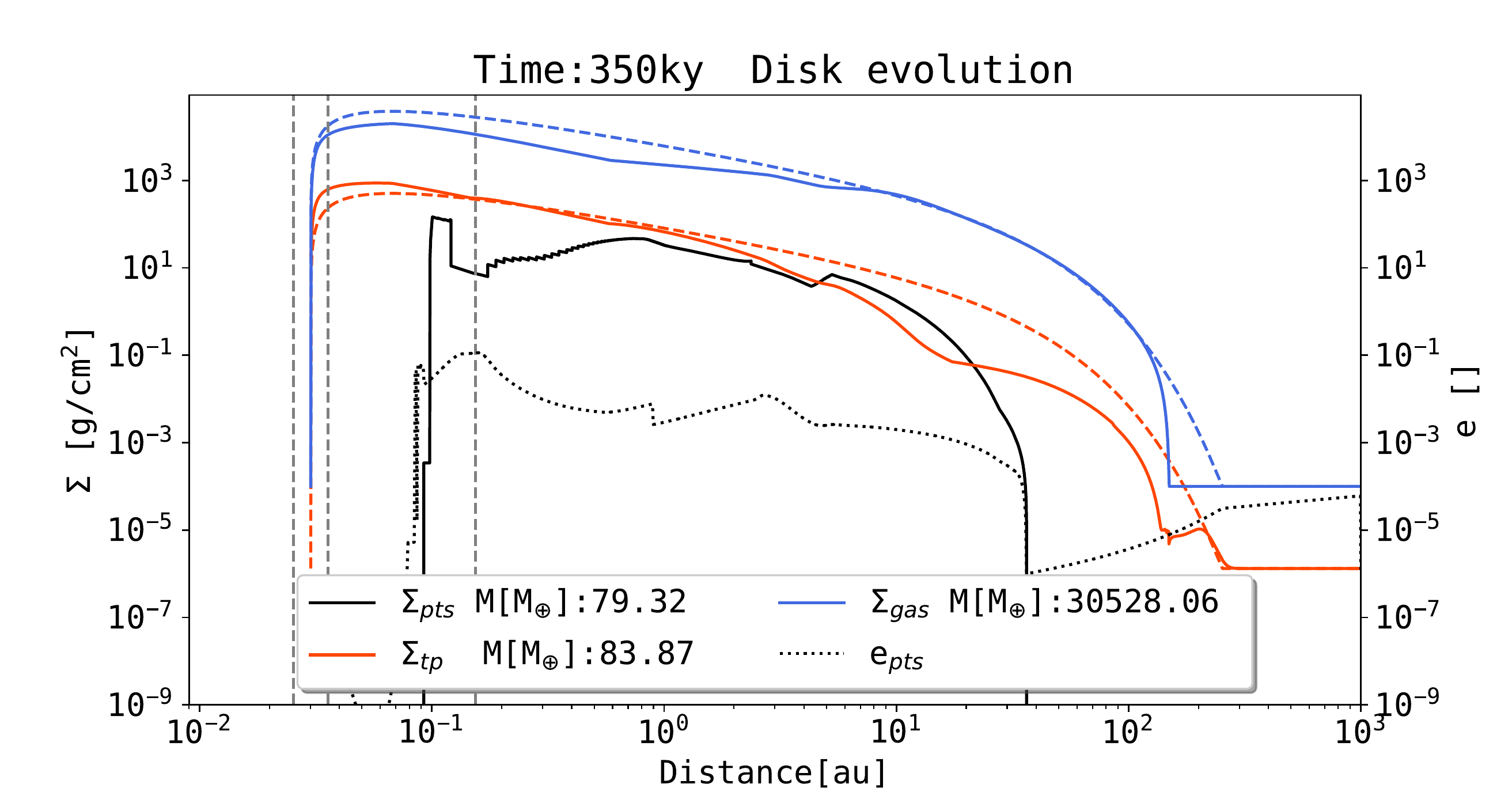} \\
\includegraphics[width=1.0\linewidth]{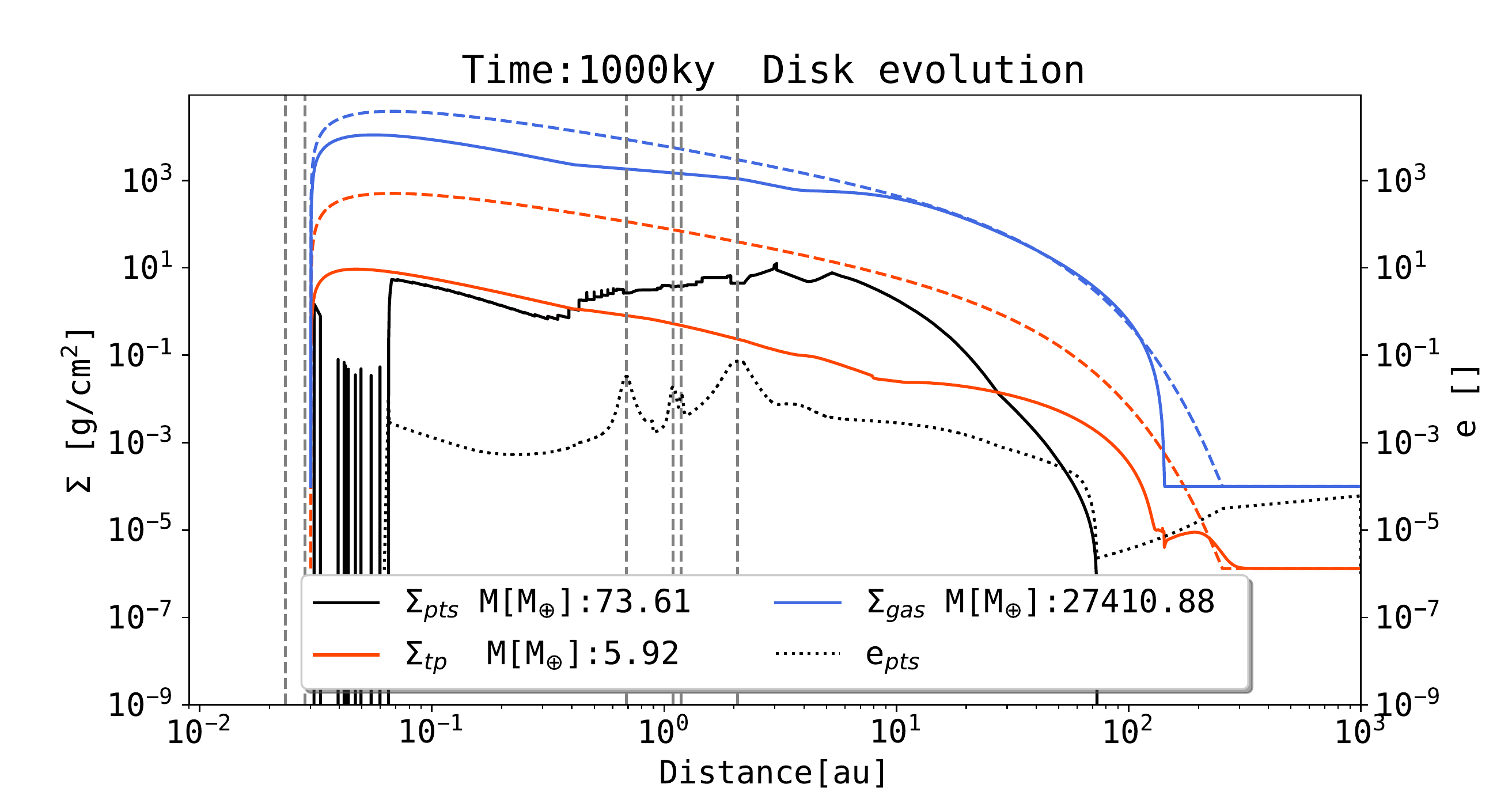} \\
\includegraphics[width=1.0\linewidth]{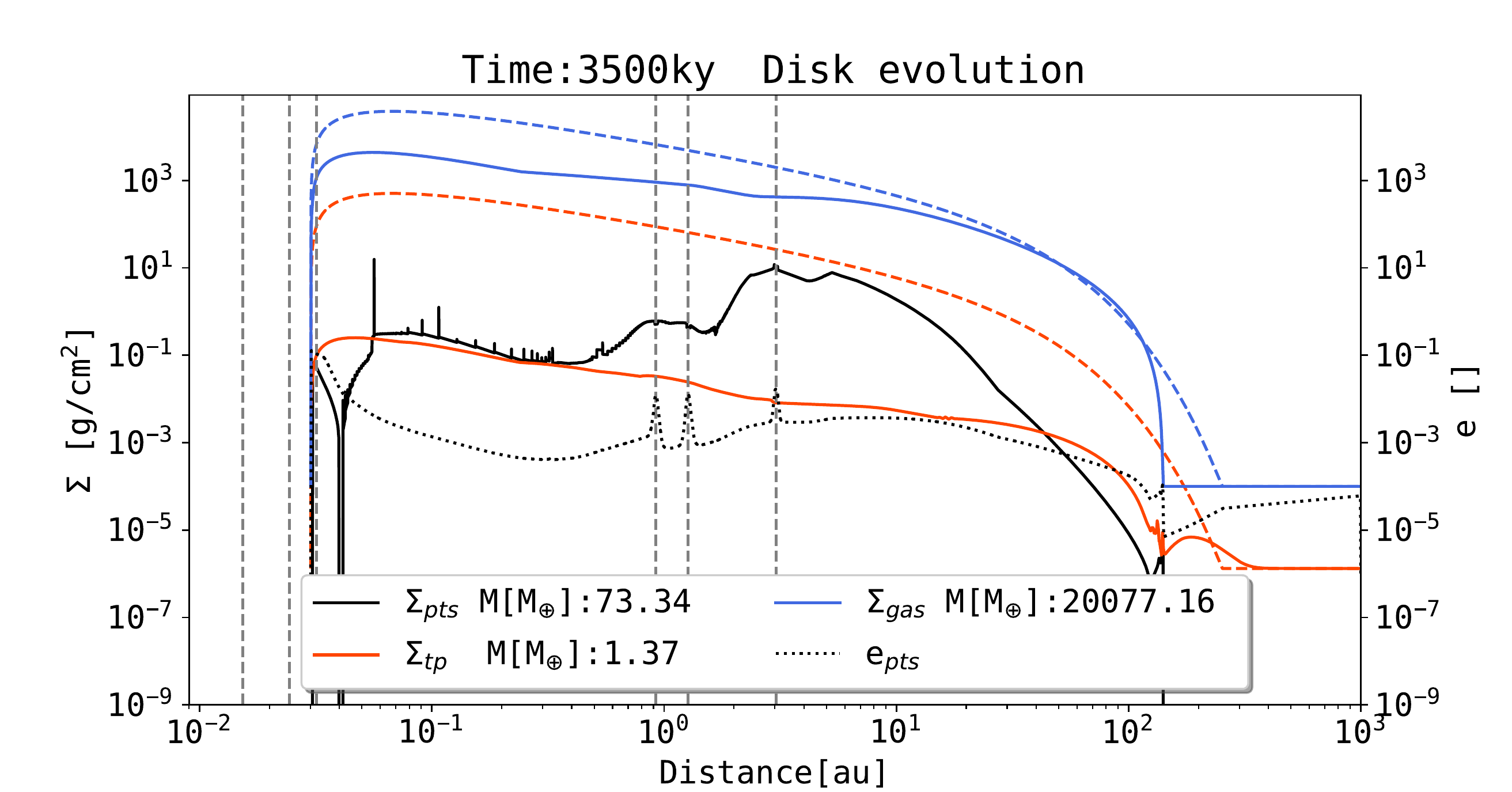} \\
\includegraphics[width=1.0\linewidth]{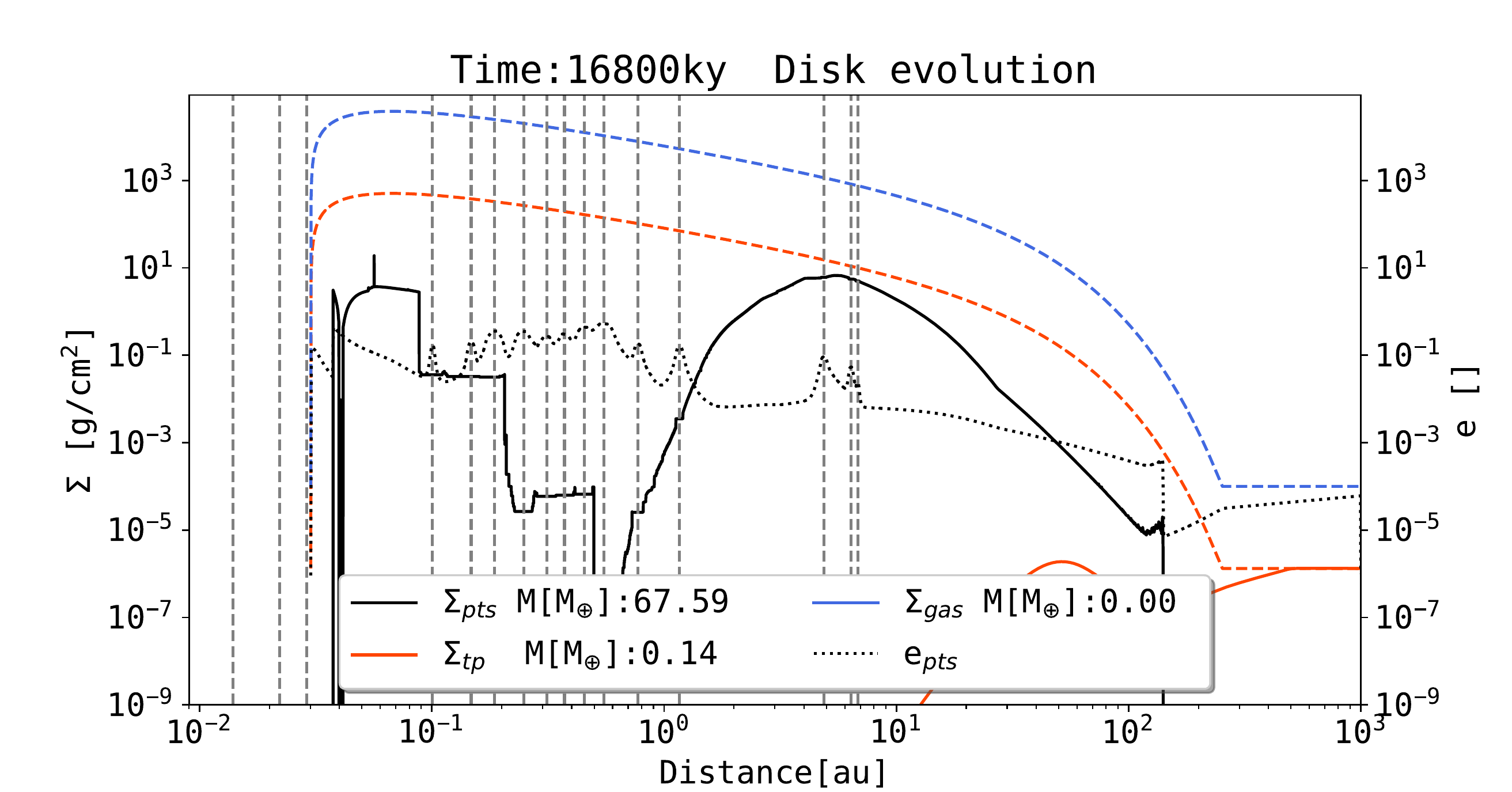} \\
\caption{\small  Surface density and planetesimal eccentricity evolution of the circumstellar disk. We show the gas surface density (blue), planetesimal surface density (black) and the combined dust and pebble surface density (orange) at t = 0.35$\,$Myr, t=1$\,$Myr, t=3.5$\,$Myr and t=16.8$\,$Myr. The initial corresponding density is shown as the dashed line. The initial mass of the gas disk is given as M$(\Sigma_{g})=34102.64\,$M$_{\oplus}$, the initial solid mass in dust and pebbles is given as M$(\Sigma_{tp})=450.93\,$M$_{\oplus}$ and the initial planetesimal disk mass is given as M$(\Sigma_{tp})=0\,$M$_{\oplus}$. The location of an active planet is indicated via dashed vertical lines. The planetesimal eccentricity is given as the dotted line respectively.
} 
\label{Fig:System_evolution_alpha_3e-4}
\end{figure}
Fig. \ref{Fig:System_evolution_alpha_3e-4} shows the surface densities and planetesimal eccentricity of the circumstellar disk at \revision{0.35$\,$Myr}, 1$\,$Myr, 3.5$\,$Myr and 16.8$\,$Myr. The semimajor axes of the active planets in the system are displayed as dashed vertical lines. Fig. \ref{Fig:System_evolution_alpha_3e-4} also shows the disk component masses at the various snapshots. 
\oliver{The very long gas disk lifetime of this setup is due to the large initial gas disk mass (0.1$\,$M$_{\odot}$) in combination with the small $\alpha = 3 \times 10^{-4}$. As discussed in \cite{Lenz_2019}, a higher photoevaporation rate would not greatly influence the formation of planetesimals, as most planetesimals form within the first Myr of the systems evolution. In order to stay consistent with \cite{Lenz_2019} we thus chose to use the same parameters. A higher photoevaporation rate to induce a shorter disk lifetime would however not affect the initial Myr of the systems evolution.}
Within the first 1$\,$Myr, the mass of the pebble and dust disk drops from an initial value of 450.93 M$_\oplus$ to only 5.92 M$_\oplus$. The mass of the planetesimal disk after 1$\,$Myr is given as 73.61 M$_\oplus$. The largest fraction of the dust and pebble disk is accreted by the host star due to continuous inward drift. The inner region of the circumstellar disk is largely depleted of planetesimals when the gas disk has vanished. After 16.8$\,$Myr, we still find 67.59$\,$M$_{\oplus}$ of planetesimals in the entire disk, most of which between 5$\,$au and 10$\,$au. 
As it can be seen in any snapshot in which planets are present, the eccentricity of planetesimals greatly increases at the location of active planets. Once the planets have migrated however, the planetesimals eccentricity is again reduced via damping by the gas disk.
 
\begin{figure}[]
\centering
\includegraphics[width=1.0\linewidth]{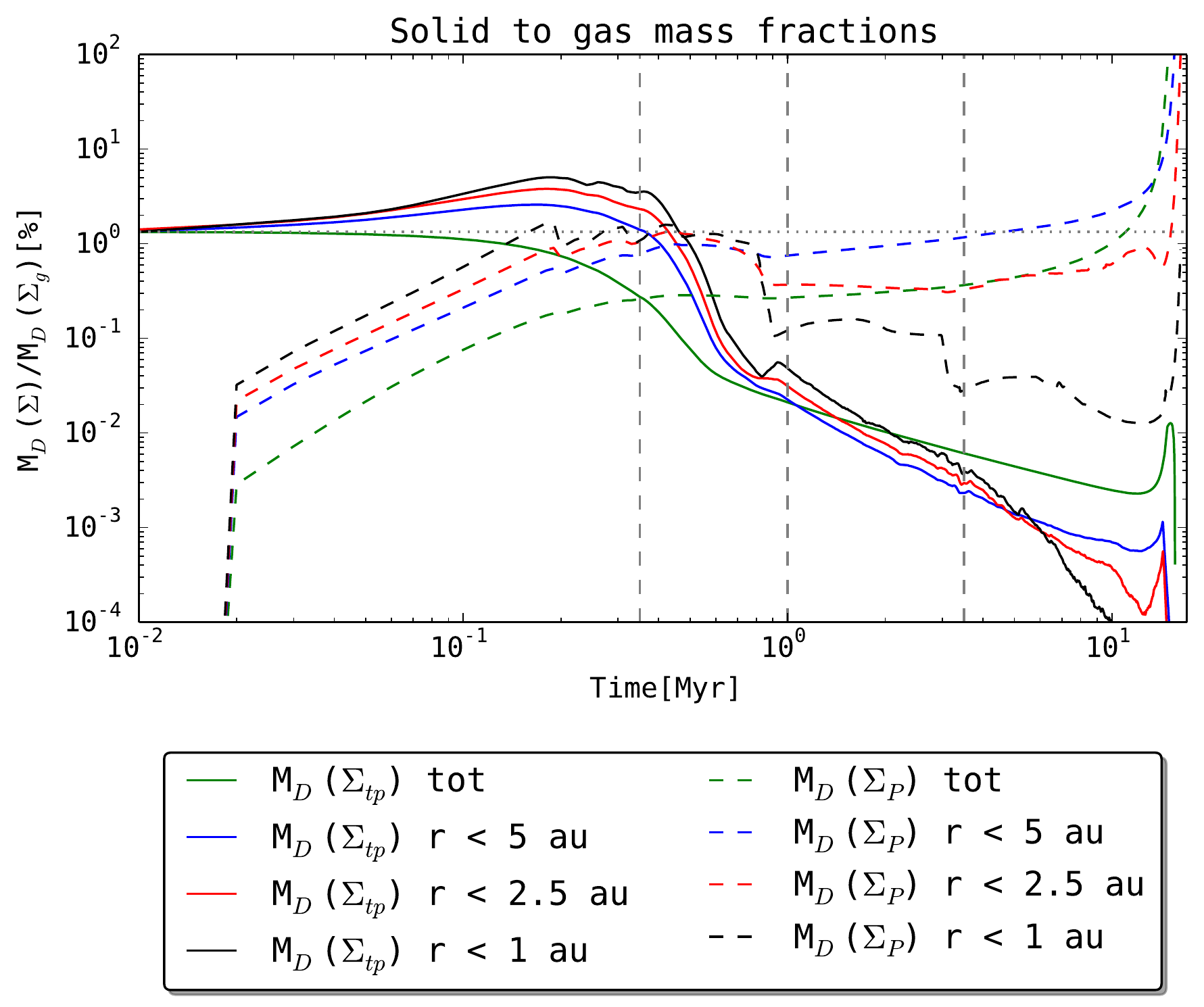} \\
\caption{\small  Fraction of the planetesimal disk mass (M$_{D}(\Sigma_{P})$) and combined dust and pebble disk mass (M$_{D}(\Sigma_{tp})$) over the gas disk mass  (M$_D({\Sigma_{g}})$) within 1$\,$au, 2.5$\,$au and 5$\,$au over time in percent. \oliver{The grey dotted line indicates the global initial dust to gas ratio $d_g$=1.34$\%$.} The dashed vertical lines are drawn at 0.35$\,$Myr, 1$\,$Myr and 3.5$\,$Myr and show the moments at which most currently active  planets are accreted by the host star or were subject to mergers. \revision{The total mass in the planetesimal disk after the dispersal of the gas disk is given as 67.59$\,$M$_{\oplus}$. The final planetesimal disk mass within 5$\,$au is given as 10.5$\,$M$_{\oplus}$, within 2.5$\,$au as 0.349$\,$M$_{\oplus}$ and within 1$\,$au as 2.35$\times 10^{-3} \,$M$_{\oplus}$.    }
} 
\label{Fig:diskMassEvolution}
\end{figure}

Fig. \ref{Fig:diskMassEvolution} shows the fraction of the planetesimal disk and the combined dust and pebble disk mass over the gas disk mass within 1$\,$au, 2.5$\,$au and 5$\,$au during the lifetime of the gas disk. As known from \cite{Lenz_2019}, the planetesimal surface density profile is steeper than the surface density profile of the gas disk, due to the influx of pebbles from distant regions of the disk. As a consequence we find that in the early phase of the evolution, the disk mass fraction of planetesimal mass over gas mass within a smaller region shows the highest value. After 200$\,$ky we find that the fraction of planetesimal mass over gas mass within 1$\,$au is >1.5$\%$, whereas the fraction of planetesimal mass over gas mass within 5$\,$au only reaches $~$0.5$\%$ at that time. If it was not for the formation of planetary embryos, we would expect this trend to continue during the lifetime of the disk. After the formation of planetary embryos however, the planetesimal disk mass within 1$\,$au strongly varies as a consequence of planetesimal accretion and continuous planetesimal formation. As can be seen in Fig. \ref{Fig:System_evolution_alpha_3e-4}, the planetesimal disk within 1$\,$au experiences the most depletion due to planets, as most embryos within 1$\,$Myr form within 1.5$\,$au. The sharp increase in the mass fraction at later times is due to the depletion of the gas disk as a consequence of photoevaporation and accretion to the host star. 
\final{The mass of the combined dust and pebble disk vastly exceeds the mass of the planetesimal disk for the first 0.35$\,$Myr within 5$\,$au, 2.5$\,$au and 1$\,$au. As the planets that form within the first 0.35$\,$Myr also formed within 2.5$\,$au, their mass growth is dominated by pebble accretion. Between 0.35$\,$Myr and 1$\,$Myr, the planetesimal disk mass exceeds the combined dust and pebble disk mass for every shown radius. As the total solid disk mass begins to be dominated by planetesimals, the accretion of pebbles is no longer the dominant mechanism of growth. Planets that form in the second generation begin to reduce their pebble mass fraction, due to the depletion of the pebble reservoir. }
\begin{comment}
\begin{figure}[]
\centering
\includegraphics[width=1.0\linewidth]{./accretedMassEvolution/alpha_3e-4_accretedMassEvolution.png} \\
%\includegraphics[width=1.0\linewidth]{./accretedMassEvolution/alpha_1e-3_accretedMassEvolution.png} \\
%\includegraphics[width=1.0\linewidth]{./accretedMassEvolution/alpha_transition_accretedMassEvolution.png} \\
\caption{\small  Evolution of the combined planet mass components for each system. We show the total pebble mass that was accreted (M$_{acc-peb}$), the total planetesimal mass that was accreted (M$_{acc-pla}$), the total envelope mass of the system (M$_{env}$) and the total mass in cores (M$_{C}$). The masses shown here do not only refer to the active planets in the system but also contain the mass components of the ones that were accreted by the star. 
} 
\label{Fig:ACCRETEDMassEvolution}
\end{figure}
\end{comment}
\subsection{Solid mass evolution}
\label{subsec:solidMassEvolution}
\begin{figure}[]
\centering
\includegraphics[width=1.0\linewidth]{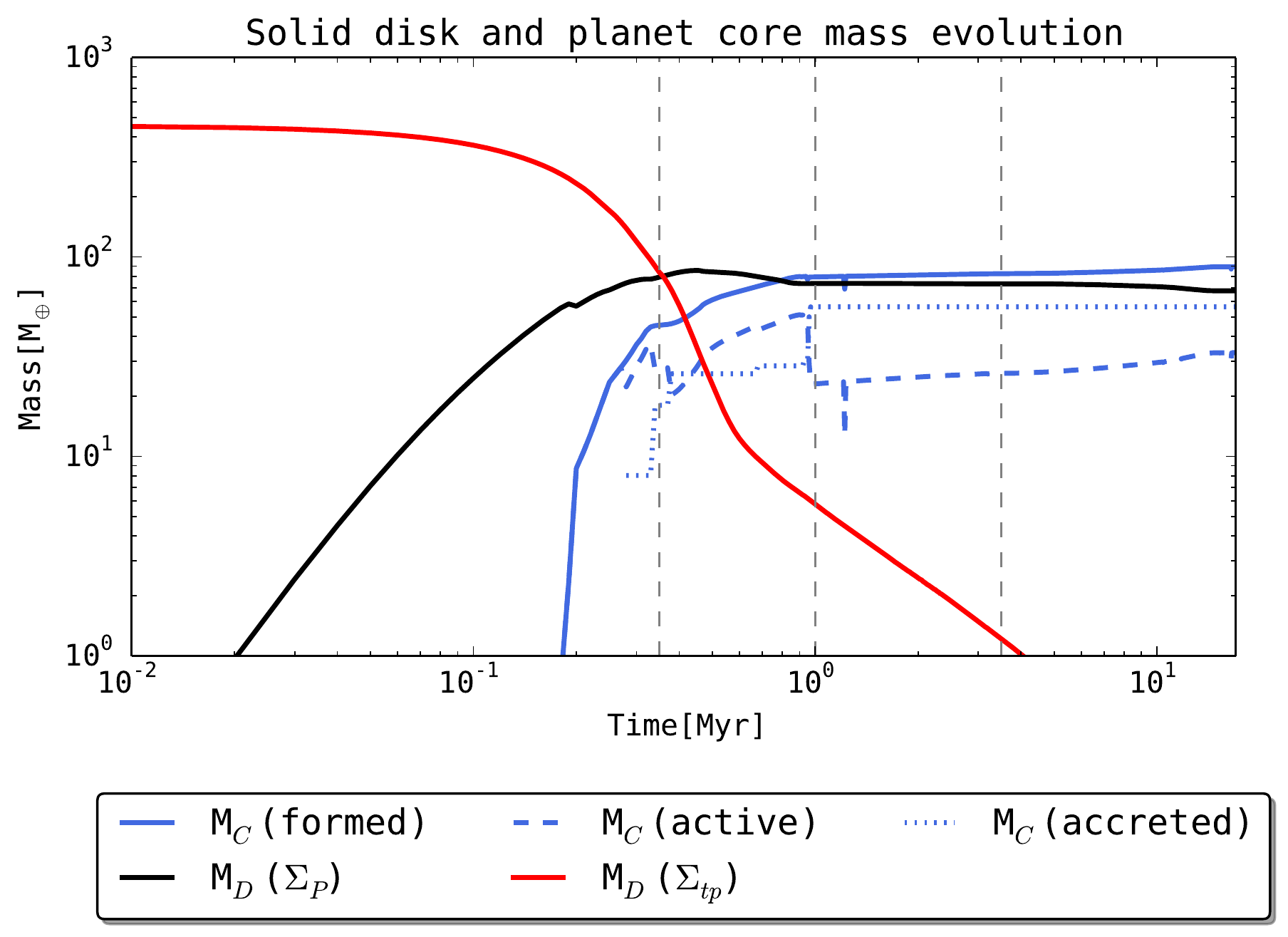}
\caption{\small  Solid mass evolution during the lifetime of the gas disk. We show the total pebble and dust disk mass (M$_D$($\Sigma_{tp}$)), the total planetesimal disk mass (M$_D$($\Sigma_{P}$)), the mass in active planetary cores (M$_C$(active)), the mass of all formed planetary cores (M$_C$(formed)) and and the mass of planetary cores that have been accreted by the host star (M$_C$(accreted)). The dashed vertical lines are drawn at 0.35$\,$Myr, 1$\,$Myr and 3.5$\,$Myr and show the moments at which most currently active  planets are accreted by the host star or were subject to mergers. 
} 
\label{Fig:solidMassEvolution}
\end{figure}

Fig. \ref{Fig:solidMassEvolution} shows the evolution of the different solid mass components of the system in Fig. \ref{Fig:semimajorAxisTime_3e-4}. This includes the total pebble and dust disk mass, the total planetesimal disk mass, the mass in all active planetary cores, the mass in all formed planetary cores and the mass of all planetary cores that were accreted by the host star. We find that the mass of cores which were accreted by the host star is larger than the remaining mass in active cores at the end of the gas disks lifetime. The mass of the planetesimal disk surpasses the mass of the dust and pebble disk after $\sim$0.35$\,$Myr. The mass of active planetary cores never surpasses the mass of the planetesimal disk, the mass of all cores that have formed however surpasses the mass of the planetesimal disk within 1$\,$Myr. The planetesimal disk mass reaches its highest value after $\sim$ 0.45$\,$Myr and then decreases. This is due to low planetesimal formation as a result of the largely depleted pebble and dust disk and planetesimal accretion onto planetary cores. 
\subsection{Final planetary system}
\label{subsec:finalSystem}
\begin{figure}[]
\centering
\includegraphics[width=1.0\linewidth]{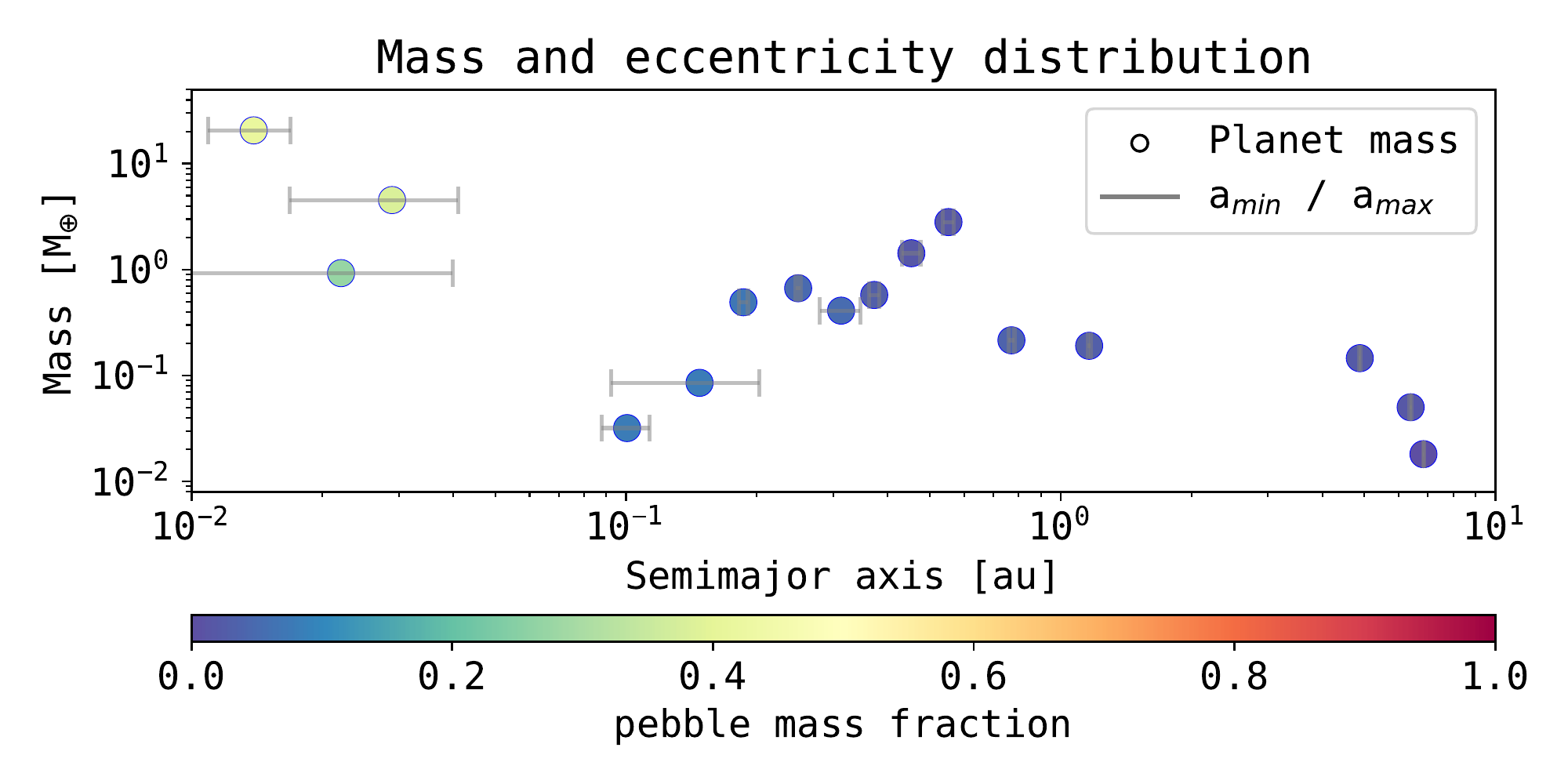} \\
\caption{\small  Mass and semimajor axis distribution of the planetary system after the gas disk has dispersed. The perihel and aphel of the planet as caused by its eccentricity is displayed via the error bars. The color shows the planets pebble mass fraction.
} 
\label{Fig:finalSystemSemiMass}
\end{figure}

Fig. \ref{Fig:finalSystemSemiMass} shows the mass, semimajor axis and eccentricity of the final system at 16.8$\,$Myr. The colormap indicates the pebble mass fraction of the individual planet. We find that the sub earth mass planets in the terrestrial planet zone are dominantly composed of planetesimals and the inner super earths show a composition that stems from both pebbles and planetesimals. There is a clear dichotomy to be found between the close in super-Earths and the planets outside of 0.1$\,$au. \revision{This dichotomy is related to orbital properties, such as the semimajor axis and individual eccentricities but also to planetary properties, such as their total mass, pebble content and embryo formation time.} The close in planets show significantly higher eccentricities than the planetesimal composed planets in the terrestrial planet zone. Most planets outside 0.1$\,$au share very low eccentricities.
 
\begin{figure}[]
\centering
\includegraphics[width=1.0\linewidth]{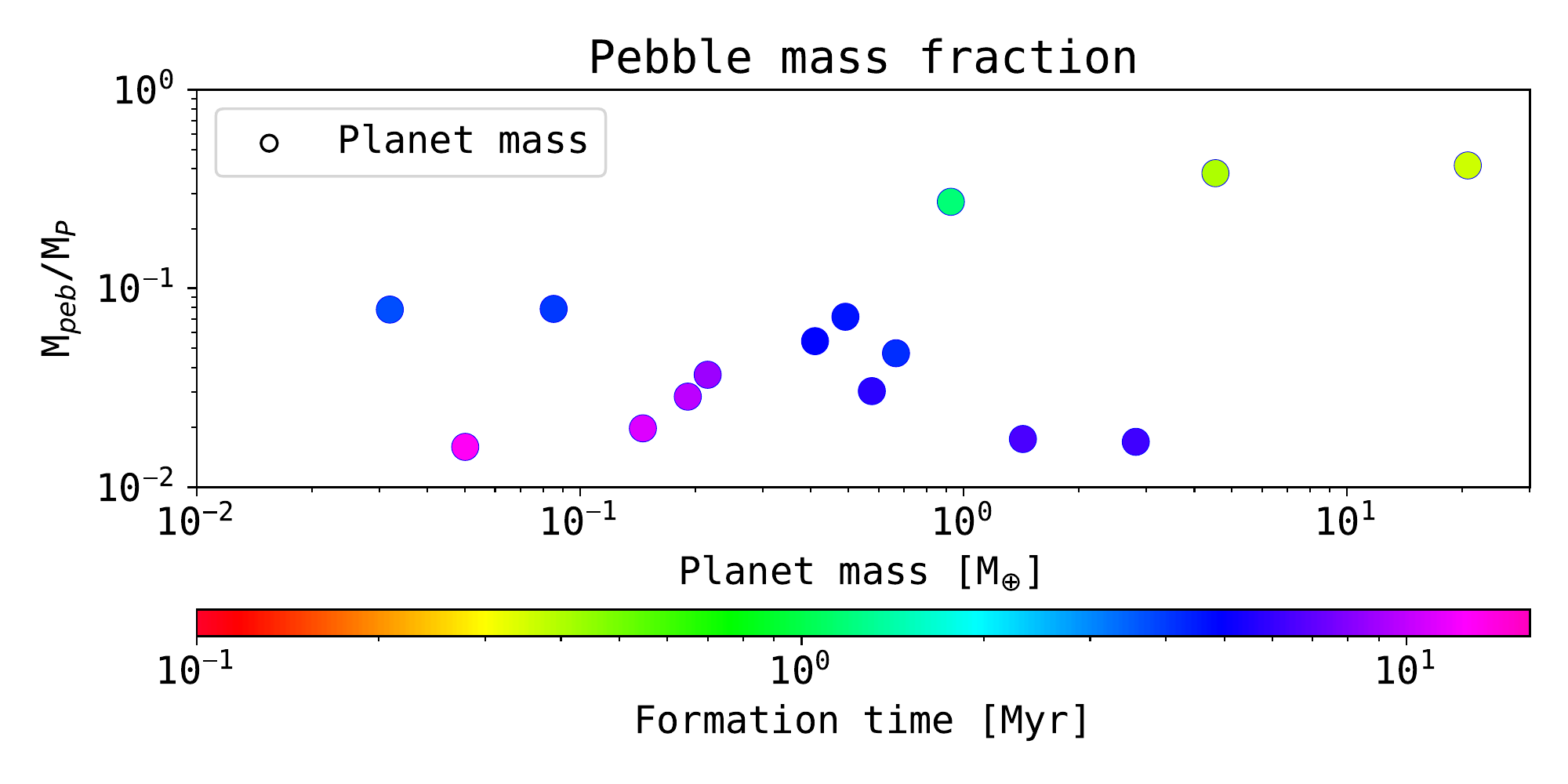} \\
\caption{\small  Mass and pebble mass fraction of the planetary system after the gas disk has dispersed. The color shows the formation time of the planets.
} 
\label{Fig:finalSystemMassPeb}
\end{figure}

Fig. \ref{Fig:finalSystemMassPeb} shows the pebble mass fraction over the planet mass for the final system after 16.8$\,$Myr. The colormap indicates the formation time of the corresponding embryo. The highest mass planets also contain the highest pebble mass fraction and the earliest formation time. The planets that formed at a later stage of the disk evolution remain at small masses and their pebble mass fraction remains below 10$\,\%$
\section{Discussion}
\subsection{The formation of multiple generations of planets}
\label{subsec:constantAlpha}
As a result of the embryo formation model, the first embryos form only in the inner region of the disk, in which they are subject to effective pebble accretion. Embryos then rapidly grow in mass and migrate to the inner edge of the disk. Since the formation of planetary embryos depends both on the planetesimal surface density and the heliocentric distance, embryos at larger distances (>10$\,$au) do not form within the lifetime of the gas disk.

As the formation of embryos in our used model does not occur within a given orbital separation to other embryos/planets, we find no embryo formation once the terrestrial planet region is populated by simultaneously growing planets. Further out embryo formation can not take place within that time, as planetesimals could not grow to a lunar mass as a consequence of larger growth time scales with orbital distances, as well as lower corresponding planetesimal surface densities. Once the super Earth mass planets have migrated to the inner edge of the gas disk and were accreted by the star, the inner region of the disk is free from planets and embryo formation from the remaining planetesimals occurs. Since there is still a larger amount of gas by the time the first super Earth planets migrated inwards, eccentricity damping of the remaining planetesimals occurs. These planetesimals were excited by the super Earth mass planets that rapidly grew to $>$10$\,$M$_{\oplus}$. The embryo growth rate depends on the dispersion velocity (see Eq. \ref{eq:mass_growth}), which again is given as $v_{\infty} = e(r) \cdot \Omega(r)$ in our framework. An increase in the eccentricity thus reduces the growth of planetary embryos. Eccentricity damping by the gas however leads again to shorter embryo growth time scales, as eccentricity damping reduces the dispersion velocity.

In our model we thus find multiple generations of planets. The first generation of planets that forms in the terrestrial planet zone grows rapidly by pebble accretion followed by rapid inward migration and subsequent accretion onto the star. Then follows a second generation with a similar fate as the first generation. The second generation of embryos also forms within the lifetime of the pebble flux. This generation grows to super earth masses via pebble accretion as well and is then subject to migration. As the pebble flux vanishes over time, the next generation cannot grow as massive as the previous ones and its migration speed is thus largely reduced. We find a set of sub earth mass planets growing by planetesimal accretion from 1$\,$Myr to 3.5$\,$Myr. Those planets however are also pushed to the host star eventually, as a more massive planet from the previous generation migrates inward as well. After the last pebble based super Earth has migrated to the inner edge of the disk after 3.5$\,$Myr, the terrestrial planet zone is free from planets once again and eccentricity damping enables the last generation to form. Until the end of the gas disk lifetime at c.a. 16.8$\,$Myr, planetary embryos can form up to a distance of 7$\,$au. The last generation grows dominantly by planetsimal accretion and remains largely at sub-Earth masses. The last generation of planets does not experience strong type I migration because of their low masses. A clearing of planets in the terrestrial planet zone as with previous generations therefore does not occur. We find that for as long as we have an active pebble flux, pebble accretion on embryos and fast type I migration clear the terrestrial zone from planets and eccentricity damping of planetesimals enables the next generation of embryos to form. 

The planets composition in terms of whether their mass stems from pebble or planetesimal accretion reflects this picture. The early generations of planets are mostly composed of pebbles, whereas the lower mass later generations of planets is dominantly formed by planetesimal accretion. The close in super Earth mass planets in the final system are composed both of pebbles and planetesimals. They stem from the second generation of planet formation during which the pebble flux largely vanished. While the most massive remaining planet initially grew mostly by pebble accretion as well, a large fraction of its mass stems from planetesimals as it continues to accrete planetesimals after the pebble flux has vanished. We also find that the highest number of planets forms in the latest generation, since type I migration no longer forces planets to the inner edge of the disk and their small masses allow for a smaller orbital spacing. 

\subsection{\revision{The relevance of pebble accretion }}
\revision{The studied system raises several questions on the role of pebble accretion for planet formation. 
Pebble accretion itself is a very efficient mechanism for planetary growth. 
But it is most efficient in the early phase of the disk evolution because it requires a pebble reservoir. 
During this phase, planetary migration is also highly efficient, due to the presence of a massive gas disk. 
The earliest planets that formed in our simulation thus experience both, rapid growth and rapid migration. 
This combination is found to be detrimental to their survival probability in our study. 
We refrain to make a general statement here on the effectiveness of pebble accretion for planet formation because our study only involves a single set of disk parameters. 
Still, we wish to highlight, that efficient growth via pebble accretion might be a destructive mechanism in planet formation for early forming embryos in the inner disk.} 

\revision{
In our study, most final planets are dominantly shaped via planetesimal accretion. 
These planets remain at lower masses because the planetesimal reservoir in the terrestrial planet zone is rather small at the time of their formation. 
The reason for this small planetesimal reservoir lies in its depletion via planetesimal accretion onto the previously formed super Earths.
While pebble accretion did not play a direct role in the formation of the last generation of planets, it does so implicitly. 
The planetesimal reservoir in the terrestrial planet zone would contain much more mass if it was not for the previous generations of super Earths that depleted it. 
The late low mass planetesimal-based generation thus required the earlier super Earth pebble-based generation to remain at such small masses and not to be carried away via migration as well.
In this scenario, pebble accretion plays a crucial role in the global process of planet formation, even though most final planets are largely composed of planetesimals.
Identifying the role of pebble accretion for a wider range of disk parameters will be subject to future work.
}
\oliver{
\subsection{Embryo formation and migration}
\label{subsec:embryoFormationAndMigration}
The model for embryo formation that is used in this study has been derived using N-body simulations including planetesimal formation \citep{voelkel2021linkingI} and has been compared to simulations that also include the effect of pebble accretion \citep{voelkel2021linkingII}. Both of these studies did not show multiple generations of planetary embryo formation within 1 Myr. Even though several setups in \cite{voelkel2021linkingII} showed the rapid growth of super Earth mass planets in the terrestrial planet zone, these super Earth mass planets did not migrate inwards, as planetary migration due to planet disk interaction was not included. The formation of a next generation was therefore suppressed due to the presence of the super Earth mass planets in the terrestrial region. Including planetary migration in the more sophisticated N-body simulations, that contain both the formation of planetesimals and the accretion of pebbles, should form multiple generations of embryos within 1 Myr as well. Such a study would support and underline the findings of this paper and will be conducted in future work.
\subsection{Long term evolution}
\label{subsec:longTermEvolution}
We chose to end our simulation after the dispersal of the gas disk because our focus lies on the dynamic embryo formation of the first couple Myr. However we still wish to briefly discuss the long term evolution of the system. After the gas disk has vanished we find 16 active planets in the system and 68.59$\,$M$_{\oplus}$ in planetesimals.
Three of those planets are very close in with masses of \revision{20.7$\,$M$_{\oplus}$ at 0.014$\,$au, 4.5$\,$M$_{\oplus}$ at 0.029$\,$au, 0.93$\,$M$_{\oplus}$ at 0.022$\,$au} and eccentric orbits. Those (super) Earth mass planets are \revision{possibly} to be accreted by the host star due to tidal interactions \revision{over timescales of Gyrs}. The remaining system would then consist of the 13 planetesimal composed planets and the remaining 68.59$\,$M$_{\oplus}$ of planetesimals. The remaining planetesimal disk mass greatly exceeds that of the 13 planetesimal composed planets, a long term integration of the system should therefore also include the remaining planetesimals and their potential embryo formation to make a concise statement on the final system after several hundred Myr. Without the damping effect of the gas disk however, the higher eccentricities of the planetesimals would reduce planetesimal accretion and embryo formation due to higher dispersion velocities.
}
\subsection{On the architecture of the solar system}
\label{subsec:architectureSolarSystem}
We wish to discuss our simulations in consideration to the initial setup of the Grand Tack model \citep{walsh2011low}. In the solar system, we find two gas giant planets at distances of 5-10$\,$au, followed by two ice giants at 19-30$\,$au. The inner region is populated with four smaller terrestrial planets. 
 As our model suggests, the last generation resembles a large set of Earth mass and sub Earth mass terrestrial embryos that are believed to form the four terrestrial planets in the Grand Tack model. The profound difference between the Grand Tack and our formation model is that the sub Earth mass terrestrial embryos did not form as a second or third generation in the Grand Tack scenario. Instead, they were merely the first generation of embryos and the reason why they did not grow to super Earth mass planets due to pebble accretion is due to Jupiter shielding the pebble flux. In contrary to the Grand Tack, our simulation suggests that a first generation of super Earth mass planets (eventually accreted by the host star) may have populated the terrestrial planet region during the first stages of the solar systems evolution.
 
As the solar system contains gas giants, which we do not form within our framework, this hypothesis is subject to further investigation. The non-formation of Jupiter in the model presented raises several profound challenges. The formation of planetary embryos is the result of planetesimals growth. This results in the very late formation of an embryo at larger distances (>5$\,$Myr at 5$\,$au). By that time, the flux of pebbles has vanished and the growth of a Jupiter core would extend the lifetime of the gas disk. The early formation of a core at a larger distance to form Jupiter would either require a single initially much larger planetesimal to form or a local overdensity in planetesimals to reduce the formation time. Such an overdensity of planetesimals is discussed to be the result of a pressure bump in the gas disk \citep{guilera2020giant}. This pressure bump however would also have major implications on the evolution of the inner system. Whether or not we also find multiple generations of terrestrial planets when a pressure bump is included will be part of future work. 
 
\final{Another possible way to form giant planets, which would require the treatment of additional physics, would be further outward migration due to orbital resonances. If two planets are captured in mean motion resonance they may form a gap in the gas disk \citep{walsh2011low}. In case the inner planet is the more massive one, this gap can cause outward migration of both planets. Effectively this might cause the formation of giant planets outside the initial orbit of their embryos formation. As gap opening in the disk is currently not included in our model, we can not observe this process within our simulation.}
 
\begin{comment}
Another pronounced feature in the solar system is the lack of planetesimals within 0.5au \ref{}. This lack of planetesimals could be explained by the lack of planetsimal formation within 0.5au, for which the reasons are found in the Stokes number of the contributing pebbles or other disk related features \ref{}. Our assumption of planetesimal formation is very generous, as all inward drifting material can contribute to planetesimal formation. An alternative theory for the lack of planetesimals within 0.5 au may be found within the multiple generations of planet formation as presented in this paper. Despite very efficient planetesimal formation within 0.5 au in our simulation, due to a very generous assumption for planetesimal formation, we find a largely depleted planetesimal disk within 0.5 au after the dispersal of the gas disk. The reason being several inward migrating super Earth mass planets that either accrete or scatter the inner planetesimals. 
\end{comment}
\section{Summary and Outlook}
In this paper we investigate the effect of dynamic planetary embryo formation during the lifetime of the gaseous disk. To pave the way for our approach we present a self consistent global model of planet formation that begins with an initial circumstellar disk of gas, dust and pebbles. The model presented viscously evolves the gas disk and uses a two population approach to model the evolution of dust and pebbles. Planetesimals form based on the radial pebble flux and planetary embryos are introduced based on the evolution of the planetesimal surface density \oliver{and their dynamical state. The eccentricities and inclinations of planetesimals are increased by nearby planetary embryos and self stirring. Simultaneously, eccentricity and inclination damping by the evolving gas disk is considered.} Once planetary embryos have formed, they can grow by pebble, planetesimal and gas accretion. Planets follow N-body dynamics with other planets and are subject to planet disk interactions, such as planetary migration. The number of embryos in the system, their initial location and formation time, are no longer an initial assumption, but the result of the disks evolution. Our main findings can be summarized as followed:
\begin{itemize}
    \item We find distinct generations of planet formation in the terrestrial planet region within the lifetime of the gas disk. Earlier generations grow dominantly by pebble accretion and are largely accreted by the host star due to migration. Later generations are composed largely of planetesimals, as those planets form after the pebble flux has mostly vanished. 
     
    \item We find close in super Earth mass planets composed of both pebbles and planetesimals and mostly planetesimal composed sub Earth mass planets in the terrestrial region. A mostly pebble composed first generation of embryos did not survive the gaseous disk, as they were accreted by the host star. The formation of close in super-Earths and mini Neptunes is a likely outcome for the early generations of planet formation.
     
    \item The majority of planetary embryos that form do not outlive the gas disk. Out of the 78 embryos that formed in total, only 16 remained after the disk has vanished. The rest is victim to either accretion to the star or mergers.
     
\end{itemize}
These findings mark the onset of a large variety of possibilities for the presented planet formation and disk evolution model. While the parameter space that we studied in this paper focused on one set of disk parameters, our model can be used in a framework of planet population synthesis as well. Additionally it can be used to study individual features of single systems in a more detailed fashion like e.g. the formation of planets in primordial rings due to pressure bumps. 
\oliver
{
Next to our presented planet formation model, we will study the possibility of multiple generations of embryo formation using large scale N-body simulations. As multiple generations appear to already form within the first 1 Myr, a sophisticated N-body study, similar to \cite{voelkel2021linkingII} is computationally feasible and should confirm our findings. The underlying hypothesis of most planet formation models states, that the final planets are the end-product from the initially placed bodies. This hypothesis is heavily challenged by our results.
}
Dynamic planetary embryo formation shows the possibility of multiple distinct generations of planet formation. This promises to have a fundamental effect on the formation history and composition of planets both in the solar system and exoplanet systems. It therefore needs to be accounted for in future studies.

\noindent
Even though we claim to start with a nebula that was designed to create the solar system \citep{lenz2020constraining}, our simulations did not lead to a planetary system that resembles our solar system. Three effects can be responsible for this: a statistical effect of the N-body solver, our systematic initial condition and/or missing physics. The statistical effect could be tested by performing numerous similar simulations and check whether this leads to a more solar system like state for a number of outcomes. MCMC simulations can then be used to further constrain the potential initial conditions that formed the solar system, like e.g. disk mass, size, profile etc. But most likely the missing physics, even without improving the turbulence model, the viscous evolution of the disk or the dust growth physics, can be crucial for not forming the solar system. A major drawback in our framework are the tidal forces acting on the disk, which are not implemented yet. Gap formation with pebble trapping or resonant outward migration of planet pairs as in the Grand Tack model can therefore not occur. This missing process however could have produced a more solar system like outcome for the chosen initial conditions. 
\section*{Acknowledgements}{We wish to thank Remo Burn and Rogerio Deienno for many fruitful discussions. This research has been supported by the Deutsche Forschungsgemeinschaft via priority program (DFG SPP) SPP 1992 "Exploring the diversity of extrasolar planets” under contract : KL 1469/17-1 Consistent Planetesimal Formation from Pebbles for Synthetic Population Syntheses of Exo-Planets and SPP 1833 "Building a Habitable Earth" under contract : KL 1469/13-1 and KL 1469/13-2.
This research was supported by the Deutsche Forschungsgemeinschaft through the Major Research Instrumentation Programme and Research Unit FOR2544 “Blue Planets around Red Stars” for H.K. under contract DFG KL1469/15-1. }
\bibliography{Template.bib} % your references Yourfile.bib

\begin{thebibliography}{68}
\expandafter\ifx\csname natexlab\endcsname\relax\def\natexlab#1{#1}\fi

\bibitem[{Adachi {et~al.}(1976)Adachi, Hayashi, \& Nakazawa}]{adachi1976gas}
Adachi, I., Hayashi, C., \& Nakazawa, K. 1976, Progress of Theoretical Physics,
  56, 1756

\bibitem[{Alibert {et~al.}(2013)Alibert, Carron, Fortier, Pfyffer, Benz,
  Mordasini, \& Swoboda}]{alibert2013theoretical}
Alibert, Y., Carron, F., Fortier, A., {et~al.} 2013, \aap, 558, A109

\bibitem[{Alibert {et~al.}(2005)Alibert, Mordasini, Benz, \&
  Winisdoerffer}]{alibert2005models}
Alibert, Y., Mordasini, C., Benz, W., \& Winisdoerffer, C. 2005, \aap, 434, 343

\bibitem[{Alibert {et~al.}(2018)Alibert, Venturini, Helled, Ataiee, Burn,
  Senecal, Benz, Mayer, Mordasini, Quanz, {et~al.}}]{alibert2018formation}
Alibert, Y., Venturini, J., Helled, R., {et~al.} 2018, Nature astronomy, 2, 873

\bibitem[{Arimatsu {et~al.}(2019)Arimatsu, Tsumura, Usui, Shinnaka, Ichikawa,
  Ootsubo, Kotani, Wada, Nagase, \& Watanabe}]{arimatsu2019kilometre}
Arimatsu, K., Tsumura, K., Usui, F., {et~al.} 2019, Nature Astronomy, 3, 301

\bibitem[{Birnstiel {et~al.}(2012)Birnstiel, Klahr, \&
  Ercolano}]{birnstiel2012simple}
Birnstiel, T., Klahr, H., \& Ercolano, B. 2012, \aap, 539, A148

\bibitem[{Bitsch \& Kley(2010)}]{bitsch2010orbital}
Bitsch, B. \& Kley, W. 2010, Astronomy \& Astrophysics, 523, A30

\bibitem[{Bitsch {et~al.}(2015)Bitsch, Lambrechts, \&
  Johansen}]{bitsch2015growth}
Bitsch, B., Lambrechts, M., \& Johansen, A. 2015, \aap, 582, A112

\bibitem[{{Bodenheimer} \& {Pollack}(1986)}]{bodenheimerpollack1986}
{Bodenheimer}, P. \& {Pollack}, J.~B. 1986, \icarus, 67, 391

\bibitem[{Bottke~Jr {et~al.}(2005)Bottke~Jr, Durda, Nesvorn{\`y}, Jedicke,
  Morbidelli, Vokrouhlick{\`y}, \& Levison}]{bottke2005linking}
Bottke~Jr, W.~F., Durda, D.~D., Nesvorn{\`y}, D., {et~al.} 2005, \icarus, 179,
  63

\bibitem[{Br{\"u}gger {et~al.}(2020)Br{\"u}gger, Burn, Coleman, Alibert, \&
  Benz}]{brugger2020pebbles}
Br{\"u}gger, N., Burn, R., Coleman, G., Alibert, Y., \& Benz, W. 2020, arXiv
  preprint arXiv:2006.04121

\bibitem[{{Chambers}(2006)}]{chambers2006}
{Chambers}, J. 2006, \icarus, 180, 496

\bibitem[{Coleman \& Nelson(2014)}]{coleman2014formation}
Coleman, G.~A. \& Nelson, R.~P. 2014, Monthly Notices of the Royal Astronomical
  Society, 445, 479

\bibitem[{Cresswell \& Nelson(2008)}]{cresswell2008three}
Cresswell, P. \& Nelson, R.~P. 2008, Astronomy \& Astrophysics, 482, 677

\bibitem[{{Crida} {et~al.}(2006){Crida}, {Morbidelli}, \& {Masset}}]{crida2006}
{Crida}, A., {Morbidelli}, A., \& {Masset}, F. 2006, \icarus, 181, 587

\bibitem[{{Delbo{\textquoteright}} {et~al.}(2017){Delbo{\textquoteright}},
  {Walsh}, {Bolin}, {Avdellidou}, \& {Morbidelli}}]{Delbo2017}
{Delbo{\textquoteright}}, M., {Walsh}, K., {Bolin}, B., {Avdellidou}, C., \&
  {Morbidelli}, A. 2017, Science, 357, 1026

\bibitem[{{Dittkrist} {et~al.}(2014){Dittkrist}, {Mordasini}, {Klahr},
  {Alibert}, \& {Henning}}]{dittkrist2014}
{Dittkrist}, K.-M., {Mordasini}, C., {Klahr}, H., {Alibert}, Y., \& {Henning},
  T. 2014, \aap, 567, A121

\bibitem[{Dittrich {et~al.}(2013)Dittrich, Klahr, \&
  Johansen}]{dittrich2013gravoturbulent}
Dittrich, K., Klahr, H., \& Johansen, A. 2013, The Astrophysical Journal, 763,
  117

\bibitem[{Emsenhuber {et~al.}(2020{\natexlab{a}})Emsenhuber, Mordasini, Burn,
  Alibert, Benz, \& Asphaug}]{emsenhuber2020new}
Emsenhuber, A., Mordasini, C., Burn, R., {et~al.} 2020{\natexlab{a}}, arXiv
  preprint arXiv:2007.05561

\bibitem[{Emsenhuber {et~al.}(2020{\natexlab{b}})Emsenhuber, Mordasini, Burn,
  Alibert, Benz, \& Asphaug}]{emsenhuber2020newpop}
Emsenhuber, A., Mordasini, C., Burn, R., {et~al.} 2020{\natexlab{b}}, arXiv
  preprint arXiv:2007.05562

\bibitem[{Fendyke \& Nelson(2014)}]{fendyke2014corotation}
Fendyke, S.~M. \& Nelson, R.~P. 2014, Monthly Notices of the Royal Astronomical
  Society, 437, 96

\bibitem[{{Fortier} {et~al.}(2013){Fortier}, {Alibert}, {Carron}, {Benz}, \&
  {Dittkrist}}]{fortier2013}
{Fortier}, A., {Alibert}, Y., {Carron}, F., {Benz}, W., \& {Dittkrist}, K.~M.
  2013, \aap, 549, A44

\bibitem[{Gonzalez(1997)}]{gonzalez1997stellar}
Gonzalez, G. 1997, Monthly Notices of the Royal Astronomical Society, 285, 403

\bibitem[{Guilera {et~al.}(2020)Guilera, S{\'a}ndor, Ronco, Venturini, \&
  Bertolami}]{guilera2020giant}
Guilera, O., S{\'a}ndor, Z., Ronco, M., Venturini, J., \& Bertolami, M. 2020,
  arXiv preprint arXiv:2005.10868

\bibitem[{Guilera {et~al.}(2010)Guilera, Brunini, \&
  Benvenuto}]{guilera2010consequences}
Guilera, O.~M., Brunini, A., \& Benvenuto, O.~G. 2010, Astronomy \&
  Astrophysics, 521, A50

\bibitem[{Harsono {et~al.}(2018)Harsono, Bjerkeli, van~der Wiel, Ramsey, Maud,
  Kristensen, \& J{\o}rgensen}]{harsono2018evidence}
Harsono, D., Bjerkeli, P., van~der Wiel, M.~H., {et~al.} 2018, Nature
  Astronomy, 2, 646

\bibitem[{Ida \& Lin(2004)}]{ida2004toward}
Ida, S. \& Lin, D.~N. 2004, \apj, 604, 388

\bibitem[{{Ida} \& {Makino}(1993)}]{idamakino1993}
{Ida}, S. \& {Makino}, J. 1993, \icarus, 106, 210

\bibitem[{Inaba {et~al.}(2001)Inaba, Tanaka, Nakazawa, Wetherill, \&
  Kokubo}]{inaba2001high}
Inaba, S., Tanaka, H., Nakazawa, K., Wetherill, G.~W., \& Kokubo, E. 2001,
  Icarus, 149, 235

\bibitem[{{Klahr} \& {Schreiber}(2020)}]{Klahr2020}
{Klahr}, H. \& {Schreiber}, A. 2020, arXiv e-prints, arXiv:2007.10696

\bibitem[{Kobayashi {et~al.}(2011)Kobayashi, Tanaka, \&
  Krivov}]{Kobayashi_2011}
Kobayashi, H., Tanaka, H., \& Krivov, A.~V. 2011, The Astrophysical Journal,
  738, 35

\bibitem[{Kokubo \& Ida(1998)}]{kokubo1998oligarchic}
Kokubo, E. \& Ida, S. 1998, Icarus, 131, 171

\bibitem[{Lambrechts {et~al.}(2014)Lambrechts, Johansen, \&
  Morbidelli}]{lambrechts2014separating}
Lambrechts, M., Johansen, A., \& Morbidelli, A. 2014, Astronomy \&
  Astrophysics, 572, A35

\bibitem[{{Lee} \& {Chiang}(2015)}]{leechiang2015}
{Lee}, E.~J. \& {Chiang}, E. 2015, \apj, 811, 41

\bibitem[{Lenz {et~al.}(2019)Lenz, Klahr, \& Birnstiel}]{Lenz_2019}
Lenz, C.~T., Klahr, H., \& Birnstiel, T. 2019, \apj, 874, 36

\bibitem[{Lenz {et~al.}(2020)Lenz, Klahr, Birnstiel, Kretke, \&
  Stammler}]{lenz2020constraining}
Lenz, C.~T., Klahr, H., Birnstiel, T., Kretke, K., \& Stammler, S. 2020,
  Astronomy \& Astrophysics, 640, A61

\bibitem[{Lin \& Papaloizou(1986)}]{lin1986tidal}
Lin, D.~N. \& Papaloizou, J. 1986, The Astrophysical Journal, 309, 846

\bibitem[{Lissauer(1993)}]{lissauer1993planet}
Lissauer, J.~J. 1993, Annual review of astronomy and astrophysics, 31, 129

\bibitem[{{L{\"u}st}(1952)}]{lust1952}
{L{\"u}st}, R. 1952, Zeitschrift Naturforschung Teil A, 7, 87

\bibitem[{{Lynden-Bell} \& {Pringle}(1974)}]{lyndenbellpringle1974}
{Lynden-Bell}, D. \& {Pringle}, J.~E. 1974, \mnras, 168, 603

\bibitem[{{Mordasini}(2018)}]{mordasini2018planetary}
{Mordasini}, C. 2018, in Handbook of Exoplanets, ed. H.~J. Deeg \& J.~A.
  Belmonte (Springer Living Reference Work), 143

\bibitem[{Mordasini {et~al.}(2009)Mordasini, Alibert, \&
  Benz}]{mordasini2009extrasolar}
Mordasini, C., Alibert, Y., \& Benz, W. 2009, \aap, 501, 1139

\bibitem[{Mordasini {et~al.}(2012)Mordasini, Alibert, Georgy, Dittkrist, Klahr,
  \& Henning}]{mordasini2012characterization}
Mordasini, C., Alibert, Y., Georgy, C., {et~al.} 2012, Astronomy \&
  Astrophysics, 547, A112

\bibitem[{{Mordasini} {et~al.}(2012){Mordasini}, {Alibert}, {Klahr}, \&
  {Henning}}]{mordasini2012combined}
{Mordasini}, C., {Alibert}, Y., {Klahr}, H., \& {Henning}, T. 2012, \aap, 547,
  A111

\bibitem[{Murray \& Chaboyer(2002)}]{murray2002stars}
Murray, N. \& Chaboyer, B. 2002, The Astrophysical Journal, 566, 442

\bibitem[{Nakamoto \& Nakagawa(1994)}]{nakamoto1994formation}
Nakamoto, T. \& Nakagawa, Y. 1994, The Astrophysical Journal, 421, 640

\bibitem[{Ndugu {et~al.}(2017)Ndugu, Bitsch, \& Jurua}]{Ndugu_2017}
Ndugu, N., Bitsch, B., \& Jurua, E. 2017, \mnras, 474, 886–897

\bibitem[{Ohtsuki {et~al.}(2002)Ohtsuki, Stewart, \&
  Ida}]{ohtsuki2002evolution}
Ohtsuki, K., Stewart, G.~R., \& Ida, S. 2002, Icarus, 155, 436

\bibitem[{Ormel \& Klahr(2010)}]{ormel2010effect}
Ormel, C. \& Klahr, H. 2010, \aap, 520, A43

\bibitem[{Ormel(2017)}]{ormel2017emerging}
Ormel, C.~W. 2017, in Formation, Evolution, and Dynamics of Young Solar Systems
  (Springer), 197--228

\bibitem[{Paardekooper {et~al.}(2011)Paardekooper, Baruteau, \&
  Kley}]{paardekooper2011torque}
Paardekooper, S.-J., Baruteau, C., \& Kley, W. 2011, Monthly Notices of the
  Royal Astronomical Society, 410, 293

\bibitem[{Picogna {et~al.}(2019)Picogna, Ercolano, Owen, \&
  Weber}]{picogna2019dispersal}
Picogna, G., Ercolano, B., Owen, J.~E., \& Weber, M.~L. 2019, Monthly Notices
  of the Royal Astronomical Society, 487, 691

\bibitem[{Pollack {et~al.}(1996)Pollack, Hubickyj, Bodenheimer, Lissauer,
  Podolak, \& Greenzweig}]{pollack1996formation}
Pollack, J.~B., Hubickyj, O., Bodenheimer, P., {et~al.} 1996, \icarus, 124, 62

\bibitem[{Rafikov(2004)}]{rafikov2004fast}
Rafikov, R.~R. 2004, The Astronomical Journal, 128, 1348

\bibitem[{Sch{\"a}fer {et~al.}(2017)Sch{\"a}fer, Yang, \&
  Johansen}]{schafer2017initial}
Sch{\"a}fer, U., Yang, C.-C., \& Johansen, A. 2017, \aap, 597, A69

\bibitem[{Schlecker {et~al.}(2021)Schlecker, Pham, Burn, Alibert, Mordasini,
  Emsenhuber, Klahr, Henning, \& Mishra}]{schlecker2021new}
Schlecker, M., Pham, D., Burn, R., {et~al.} 2021, arXiv preprint
  arXiv:2104.11750

\bibitem[{Schlichting {et~al.}(2013)Schlichting, Fuentes, \&
  Trilling}]{schlichting2013initial}
Schlichting, H.~E., Fuentes, C.~I., \& Trilling, D.~E. 2013, \aj, 146, 36

\bibitem[{Seager {et~al.}(2007)Seager, Kuchner, Hier-Majumder, \&
  Militzer}]{seager2007mass}
Seager, S., Kuchner, M., Hier-Majumder, C., \& Militzer, B. 2007, The
  Astrophysical Journal, 669, 1279

\bibitem[{Shakura \& Sunyaev(1973)}]{shakura1973black}
Shakura, N.~I. \& Sunyaev, R.~A. 1973, \aap, 24, 337

\bibitem[{{Thommes} {et~al.}(2003){Thommes}, {Duncan}, \&
  {Levison}}]{thommes2003}
{Thommes}, E.~W., {Duncan}, M.~J., \& {Levison}, H.~F. 2003, \icarus, 161, 431

\bibitem[{Voelkel {et~al.}(2021{\natexlab{a}})Voelkel, Deienno, Kretke, \&
  Klahr}]{voelkel2021linkingI}
Voelkel, O., Deienno, R., Kretke, K., \& Klahr, H. 2021{\natexlab{a}},
  Astronomy \& Astrophysics, 645, A131

\bibitem[{Voelkel {et~al.}(2021{\natexlab{b}})Voelkel, Deienno, Kretke, \&
  Klahr}]{voelkel2021linkingII}
Voelkel, O., Deienno, R., Kretke, K., \& Klahr, H. 2021{\natexlab{b}},
  Astronomy \& Astrophysics, 645, A132

\bibitem[{Voelkel {et~al.}(2020)Voelkel, Klahr, Mordasini, Emsenhuber, \&
  Lenz}]{voelkel2020effect}
Voelkel, O., Klahr, H., Mordasini, C., Emsenhuber, A., \& Lenz, C. 2020,
  Astronomy \& Astrophysics, 642, A75

\bibitem[{Walsh {et~al.}(2017)Walsh, Bolin, Avdellidou, Morbidelli,
  {et~al.}}]{walsh2017identification}
Walsh, K., Bolin, B., Avdellidou, C., Morbidelli, A., {et~al.} 2017, Science,
  357, 1026

\bibitem[{Walsh \& Levison(2019)}]{walsh2019planetesimals}
Walsh, K.~J. \& Levison, H.~F. 2019, Icarus, 329, 88

\bibitem[{Walsh {et~al.}(2011)Walsh, Morbidelli, Raymond, O'Brien, \&
  Mandell}]{walsh2011low}
Walsh, K.~J., Morbidelli, A., Raymond, S.~N., O'Brien, D.~P., \& Mandell, A.~M.
  2011, Nature, 475, 206

\bibitem[{Weidenschilling(2011)}]{weidenschilling2011initial}
Weidenschilling, S. 2011, \icarus, 214, 671

\bibitem[{Zheng {et~al.}(2017)Zheng, Lin, \&
  Kouwenhoven}]{zheng2017planetesimal}
Zheng, X., Lin, D.~N., \& Kouwenhoven, M. 2017, The Astrophysical Journal, 836,
  207

\end{thebibliography}
\end{document}